\PassOptionsToPackage{dvipsnames}{xcolor}
\documentclass[preprint]{aastex631}
\usepackage{amsmath}
\usepackage{bm}
\usepackage{capt-of}   
\usepackage{svg}
\usepackage{empheq}
\usepackage{enumitem}
\usepackage{hyperref}
\usepackage{hhline}
\usepackage{longtable}
\usepackage{numprint}
\usepackage{pifont}
\usepackage{xspace}
\usepackage{placeins}
\usepackage{graphicx}
\usepackage{mwe}
\usepackage[bottom]{footmisc}

\newcommand{\nparams}{41\xspace}

\newcommand{\rf}{\textcolor{black}}

\newcommand{\ed}{\textcolor{black}}

\newcommand{\ang}{\AA\xspace}

\newcommand{\projname}{DESI Strong Lens Foundry\xspace}

\newcommand{\ho}{\ensuremath{H_0}\xspace}

\newcommand{\gigal}{\texttt{GIGA-Lens}\xspace}
\newcommand{\thetaEfit}{\ensuremath{2.646\pm0.002\twopr}\xspace}
\newcommand{\gammafit}{\ensuremath{1.37\pm0.02}\xspace}
\newcommand{\Refit}{\ensuremath{1.89\pm0.02\twopr}\xspace}
\newcommand{\Refitkpc}{\ensuremath{11.8\pm0.1}~kpc\xspace}

\newcommand{\rhat}{\ensuremath{\hat{R}}\xspace}
\newcommand{\tE}{\ensuremath{\theta_E}\xspace}

\newcommand{\exetime}{{{3~\rm{hrs.} 13~\rm{min.}}}\xspace}

\newcommand{\HST}{\emph{Hubble Space Telescope}\xspace}
\newcommand{\hst}{\emph{HST}\xspace}
\newcommand{\JWST}{\emph{James Webb Space Telescope}\xspace}

\newcommand{\RST}{\emph{Nancy Grace Roman Space Telescope}\xspace}

\newcommand\pound{\scalebox{0.8}{\raisebox{0.4ex}{\#}}}

\newcommand{\coverage}{19,000\xspace}

\newcommand{\ser}{S{\'e}rsic\xspace}

\newcommand{\twopr}{^{\prime \prime}}

\submitjournal{ApJ}

\shorttitle{\projname I}
\shortauthors{Huang, Baltasar, Ratier-Werbin et al.}

\begin{document}
\title{\projname I:\\
\hst Observations and Modeling with \gigal}


\correspondingauthor{Xiaosheng Huang}
\email{xhuang22@usfca.edu}

\author[0000-0001-8156-0330]{X.~Huang}
\affiliation{Department of Physics \& Astronomy, University of San Francisco, San Francisco, CA 94117, USA}
\affiliation{Physics Division, Lawrence Berkeley National Laboratory, 1 Cyclotron Road, Berkeley, CA 94720, USA}

\author[0009-0003-4697-7079]{S.~Baltasar}
\affiliation{Physics Division, Lawrence Berkeley National Laboratory, 1 Cyclotron Road, Berkeley, CA 94720, USA}
\affiliation{Department of Physics, 
Complutense University of Madrid, 28040 Madrid, Spain}
\affiliation{Department of Mathematics, 
Complutense University of Madrid, 28040 Madrid, Spain}

\author[0009-0009-8206-0325]{N.~Ratier-Werbin}
\affiliation{Physics Division, Lawrence Berkeley National Laboratory, 
1 Cyclotron Road, Berkeley, CA 94720, USA}
\affiliation{Department of Physics, 
Complutense University of Madrid, 28040 Madrid, Spain}
\affiliation{Department of Mathematics, 
Complutense University of Madrid, 28040 Madrid, Spain}

\author[0000-0002-0385-0014]{C.~Storfer}
\affiliation{Institute for Astronomy, University of Hawai'i, 
Honolulu, HI 96822-1897, USA}

\author[0000-0003-1889-0227]{W.~Sheu}
\affiliation{Department of Physics \& Astronomy, 
University of California, Los Angeles, Los Angeles, CA 90095, USA}
\affiliation{Physics Division, Lawrence Berkeley National Laboratory, 
1 Cyclotron Road, Berkeley, CA 94720, USA}

\author[0000-0002-2350-4610]{S.~Agarwal}
\affiliation{University of Chicago, Department of Astronomy, Chicago, IL 60615, USA}
\affiliation{Physics Division, Lawrence Berkeley National Laboratory, 1 Cyclotron Road, Berkeley, CA 94720, USA}

\author[0009-0008-0518-8045]{M.~Tamargo-Arizmendi}
\affiliation{Physics Division, Lawrence Berkeley National Laboratory, 
1 Cyclotron Road, Berkeley, CA 94720, USA}
\affiliation{Department of Physics \& Astronomy, University of Pittsburgh, Pittsburgh, PA 15260, USA}

\author[0000-0002-5042-5088]{D.J.~Schlegel}
\affiliation{Physics Division, Lawrence Berkeley National Laboratory, 
1 Cyclotron Road, Berkeley, CA 94720, USA}

\author{J.~Aguilar}
\affiliation{Physics Division, Lawrence Berkeley National Laboratory, 1 Cyclotron Road, Berkeley, CA 94720, USA}

\author[0000-0001-6098-7247]{S.~Ahlen}
\affiliation{Physics Dept., Boston University, 590 Commonwealth Avenue, Boston, MA 02215, USA}

\author{G.~Aldering}
\affiliation{Physics Division, Lawrence Berkeley National Laboratory, 1 Cyclotron Road, Berkeley, CA 94720, USA}

\author{S.~Banka}
\affiliation{Department of Electrical Engineering \& Computer Sciences, 
University of California, Berkeley, Berkeley, CA 94720}

\author[0000-0001-5537-4710]{S.~BenZvi}
\affiliation{Department of Physics \& Astronomy, University of Rochester, 
206 Bausch and Lomb Hall, P.O. Box 270171, 
Rochester, NY 14627-0171, USA}

\author[0000-0001-9712-0006]{D.~Bianchi}
\affiliation{Dipartimento di Fisica ``Aldo Pontremoli'', Universit\`a degli Studi di Milano, 
Via Celoria 16, I-20133 Milano, Italy}

\author[0000-0002-9836-603X]{A.~Bolton}
\affiliation{NSF's National Optical-Infrared Astronomy Research Laboratory, Tucson, AZ 85719, USA}

\author{D.~Brooks}
\affiliation{Department of Physics \& Astronomy, University College London, 
Gower Street, London, WC1E 6BT, UK}

\author[0000-0001-7101-9831]{A.~Cikota}
\affiliation{Gemini Observatory / NSF's NOIRLab, Casilla 603, La Serena, Chile}

\author{T.~Claybaugh}
\affiliation{Physics Division, Lawrence Berkeley National Laboratory, 
1 Cyclotron Road, Berkeley, CA 94720, USA}

\author[0000-0002-1769-1640]{A.~de~la~Macorra}
\affiliation{Instituto de F\'{\i}sica, Universidad Nacional Aut\'{o}noma de M\'{e}xico,  
Circuito de la Investigaci\'{o}n Cient\'{\i}fica, Ciudad Universitaria, 
Cd. de M\'{e}xico  C.~P.~04510,  M\'{e}xico}

\author[0000-0002-4928-4003]{A.~Dey}
\affiliation{NSF's National Optical-Infrared Astronomy Research Laboratory, Tucson, AZ 85719, USA}

\author{P.~Doel}
\affiliation{Department of Physics \& Astronomy, University College London, 
Gower Street, London, WC1E 6BT, UK}

\author{J.~Edelstein}
\affiliation{Space Sciences Laboratory, University of California, Berkeley, 7 Gauss Way, Berkeley, CA 94720, USA}

\author[0000-0003-4701-3469]{A.~Filipp}
\affiliation{Universit\'{e} de Montréal, Physics Department, 
1375 Av. Th\'{e}r\`{e}se-Lavoie-Roux, H2V~0B3 Montr\'{e}al, QC, Canada}
\affiliation{Ciela -- Montreal Institute for Astrophysical Data Analysis and Machine Learning, 
1375 Av. Th\'{e}r\`{e}se-Lavoie-Roux, H2V~0B3 Montr\'{e}al, QC, Canada}
\affiliation{Technical University Munich, TUM School of Natural Sciences, 
Physics Department, 85748 Garching, Germany}
\affiliation{Max Planck Institute for Astrophysics (MPA), Karl-Schwarzschild-Strasse 1, 
85748 Garching, Germany}

\author[0000-0002-2890-3725]{J.~E.~Forero-Romero}
\affiliation{Departamento de F\'isica, Universidad de los Andes, Cra. 1 No. 18A-10, Edificio Ip, CP 111711, Bogot\'a, Colombia}
\affiliation{Observatorio Astron\'omico, Universidad de los Andes, 
Cra. 1 No. 18A-10, Edificio H, CP 111711 Bogot\'a, Colombia}

\author{E.~Gazta\~{n}aga}
\affiliation{Institut d'Estudis Espacials de Catalunya (IEEC), 
c/ Esteve Terradas 1, Edifici RDIT, Campus PMT-UPC, 08860 Castelldefels, Spain}
\affiliation{Institute of Cosmology and Gravitation, 
University of Portsmouth, Dennis Sciama Building, Portsmouth, PO1 3FX, UK}
\affiliation{Institute of Space Sciences, ICE-CSIC, Campus UAB, 
Carrer de Can Magrans s/n, 08913 Bellaterra, Barcelona, Spain}

\author[0000-0003-3142-233X]{S.~Gontcho~A~Gontcho}
\affiliation{Physics Division, Lawrence Berkeley National Laboratory, 
1 Cyclotron Road, Berkeley, CA 94720, USA}

\author[0000-0003-2748-7333]{A.~Gu}
\affiliation{Quantum Science and Engineering, 
Harvard University, Cambridge, MA 02138, USA}

\author{G.~Gutierrez}
\affiliation{Fermi National Accelerator Laboratory, PO Box 500, Batavia, IL 60510, USA}

\author[0000-0002-6550-2023]{K.~Honscheid}
\affiliation{Center for Cosmology and AstroParticle Physics, 
The Ohio State University, 191 West Woodruff Avenue, Columbus, OH 43210, USA}
\affiliation{Department of Physics, The Ohio State University, 
191 West Woodruff Avenue, Columbus, OH 43210, USA}
\affiliation{The Ohio State University, Columbus, 43210 OH, USA}


\author[0000-0002-9253-053X]{E.~Jullo}
\affiliation{Aix-Marseille Univ., CNRS, CNES, LAM, Marseille, France}

\author[0000-0002-0000-2394]{S.~Juneau}
\affiliation{NSF NOIRLab, 950 N. Cherry Ave., Tucson, AZ 85719, USA}

\author{R.~Kehoe}
\affiliation{Department of Physics, Southern Methodist University, 
3215 Daniel Avenue, Dallas, TX 75275, USA}

\author[0000-0002-8828-5463]{D.~Kirkby}
\affiliation{Department of Physics and Astronomy, 
University of California, Irvine, CA 92697, USA}

\author[0000-0003-3510-7134]{T.~Kisner}
\affiliation{Physics Division, Lawrence Berkeley National Laboratory, 
1 Cyclotron Road, Berkeley, CA 94720, USA}

\author[0000-0001-6356-7424]{A.~Kremin}
\affiliation{Physics Division, Lawrence Berkeley National Laboratory, 
1 Cyclotron Road, Berkeley, CA 94720, USA}

\author[0000-0001-9802-362X]{K.J.~Kwon}
\affiliation{Department of Physics, University of California, Santa Barbara, 
Santa Barbara, CA 93106, USA}

\author{A.~Lambert}
\affiliation{Physics Division, Lawrence Berkeley National Laboratory, 
1 Cyclotron Road, Berkeley, CA 94720, USA}

\author[0000-0003-1838-8528]{M.~Landriau}
\affiliation{Physics Division, Lawrence Berkeley National Laboratory, 
1 Cyclotron Road, Berkeley, CA 94720, USA}

\author[0000-0002-1172-0754]{D.~Lang}
\affiliation{Perimeter Institute for Theoretical Physics, 
Waterloo, ON N2L 2Y5, Canada}

\author[0000-0001-7178-8868]{L.~Le~Guillou}
\affiliation{Sorbonne Universit\'{e}, CNRS/IN2P3, 
Laboratoire de Physique Nucl\'{e}aire et de Hautes Energies (LPNHE), 
FR-75005 Paris, France}

\author{J.~Liu}
\affiliation{Department of Physics, Rose-Hulman Institute of Technology, 
Terre Haute, IN 47803, USA}

\author[0000-0002-1125-7384]{A.~Meisner}
\affiliation{NSF's National Optical-Infrared Astronomy Research Laboratory, Tucson, AZ 85719, USA}

\author{R.~Miquel}
\affiliation{Instituci\'{o} Catalana de Recerca i Estudis Avan\c{c}ats, 
Passeig de Llu\'{\i}s Companys, 23, 08010 Barcelona, Spain}
\affiliation{Institut de F\'{i}sica d’Altes Energies (IFAE), 
The Barcelona Institute of Science and Technology, Edifici Cn, Campus UAB, 
08193, Bellaterra (Barcelona), Spain}

\author[0000-0002-2733-4559]{J.~Moustakas}
\affiliation{Department of Physics and Astronomy, 
Siena College, 515 Loudon Road, Loudonville, NY 12211, USA}

\author{A.~D.~Myers}
\affiliation{Department of Physics \& Astronomy, 
University of Wyoming, 1000 E. University, Dept.~3905, 
Laramie, WY 82071, USA}

\author[0000-0002-4436-4661]{S.~Perlmutter}
\affiliation{Physics Division, Lawrence Berkeley National Laboratory, 
1 Cyclotron Road, Berkeley, CA 94720, USA}
\affiliation{Department of Physics, University of California, 
Berkeley, CA 94720, USA}

\author[0000-0001-6979-0125]{I.~P\'erez-R\`afols}
\affiliation{Departament de F\'isica, EEBE, Universitat Polit\`ecnica de Catalunya, 
c/Eduard Maristany 10, 08930 Barcelona, Spain}

\author[0000-0001-7145-8674]{F.~Prada}
\affiliation{Instituto de Astrof\'{i}sica de Andaluc\'{i}a (CSIC), 
Glorieta de la Astronom\'{i}a, s/n, E-18008 Granada, Spain}

\author{G.~Rossi}
\affiliation{Department of Physics and Astronomy, 
Sejong University, 209 Neungdong-ro, Gwangjin-gu, Seoul 05006, Republic of Korea}

\author[0000-0001-5402-4647]{D.~Rubin}
\affiliation{Department of Physics \& Astronomy, 
University of Hawai'i at M\~{a}noa, Honolulu, HI 96822, USA}
\affiliation{Physics Division, Lawrence Berkeley National Laboratory, 
1 Cyclotron Road, Berkeley, CA 94720, USA}

\author[0000-0002-9646-8198]{E.~Sanchez}
\affiliation{CIEMAT, Avenida Complutense 40, E-28040 Madrid, Spain}

\author{M.~Schubnell}
\affiliation{Department of Physics, University of Michigan, 
450 Church Street, Ann Arbor, MI 48109, USA}
\affiliation{University of Michigan, 500 S. State Street, 
Ann Arbor, MI 48109, USA}

\author{Y.~Shu}
\affiliation{Purple Mountain Observatory, Chinese Academy of Sciences, 
Nanjing 210023, People's Republic of China}

\author{E.~Silver}
\affiliation{Department of Astronomy, University of California, 
Berkeley, CA 94720, USA}

\author{D.~Sprayberry}
\affiliation{NSF's National Optical-Infrared Astronomy Research Laboratory, Tucson, AZ 85719, USA}

\author[0000-0001-7266-930X]{N.~Suzuki}
\affiliation{Physics Division, Lawrence Berkeley National Laboratory, 
1 Cyclotron Road, Berkeley, CA 94720, USA}


\author[0000-0003-1704-0781]{G.~Tarl\'{e}}
\affiliation{University of Michigan, 500 S. State Street, 
Ann Arbor, MI 48109, USA}

\author{B.A.~Weaver}
\affiliation{NSF's National Optical-Infrared Astronomy Research Laboratory, Tucson, AZ 85719, USA}

\author[0000-0002-6684-3997]{H.~Zou}
\affiliation{National Astronomical Observatories, Chinese Academy of Sciences, 
A20 Datun Rd., Chaoyang District, Beijing, 100012, P.R. China}







\begin{abstract}
We present the Dark Energy Spectroscopic Instrument (DESI) Strong Lens Foundry. We discovered $\sim 3500$ new strong gravitational lens candidates in the DESI Legacy Imaging Surveys
using residual neural networks (ResNet). We observed a subset (51) of our candidates using the {\it Hubble Space Telescope} ({\it HST}). All of them were confirmed to be strong lenses. We also briefly describe spectroscopic follow-up observations by DESI and Keck NIRES programs. From this very rich dataset, a number of studies will be carried out, including evaluating the quality of the ResNet search candidates and lens modeling. In this paper, we present our initial effort in these directions. 
In particular, as a demonstration, we present the lens model for DESI-165.4754-06.0423, with imaging data from {\it HST}, and lens and source redshifts from DESI and Keck NIRES, respectively. 
In this effort, we have applied a \emph{fully} forward-modeling Bayesian approach (\texttt{GIGA-Lens}), using \emph{multiple} GPUs, for the first time in both regards, to a strong lens with {\it HST} data, or any high resolution imaging.
\end{abstract}
\keywords{galaxies: high-redshift -- gravitational lensing: strong 
}

\section{Introduction}\label{sec:intro}
Strong gravitational lensing systems 
are a powerful tool for astrophysics and cosmology.
They have been used to study how dark matter is distributed in galaxies and \rf{galaxy} clusters 
\citep[e.g.,][]{kochanek1991a, blandford1992a, broadhurst2000a,  koopmans2002a, 
bolton2006a, clowe2006a, koopmans2006a, bradac2008a, vegetti2009a, huang2009a, jullo2010a,
tessore2016a, monna2017a, jauzac2018a, 
shajib2019a, meneghetti2020a}.
Furthermore, carefully modeling the mass profiles of galaxy-scale strong lenses for a large number of lensing systems 
over a wide range of redshifts makes it possible to  study the structural evolution 
of massive elliptical galaxies \ed{\citep[e.g., ][]{bolton2012b, sonnenfeld2013b, LiR2018a, filipp2023a}}.
They are also uniquely suited to probe 
dark matter substructure beyond the local universe and line-of-sight low-mass halos
\citep[e.g.,][]{vegetti2014a, hezaveh2016a, vegetti2018a, ritondale2019a, diazrivero2020a, cagan-sengul2022a, nierenberg2023a}, 
to test the predictions of the cold dark matter (CDM) model.

Recent measurements of the Hubble constant \ho span a range of $\sim$10\% \citep[e.g.,][]{abbott2017a, abbott2018b, riess2019a, wong2019a, freedman2019a, freedman2020a, planck2020a, khetan2020a, philcox2020a, choi2020a, dhawan2023a}, 
and significant tension between predictions for \ho based on early-universe observables 
and direct late-universe measurements remain \citep[e.g.,][]{verde2019a}.
Multiply-lensed supernovae (SNe) are 
ideal for measuring time-delays and \ho because of their well-characterized light curves,
and in the case of Type~Ia, with the added benefit of standardizable luminosity \citep{refsdal1964a, treu2010a, oguri2010a, suyu2023a}.
In recent years, strongly lensed supernovae, 
both core-collapse \citep{kelly2015a, rodney2016a} and Type Ia \citep{quimby2014a, goobar2017a, rodney2021a, chen2022a, goobar2023a, pierel2023a, frye2023a}, have been discovered.
Retrospective searches have found lensed SNe as well, \ed{including} Type Ia's \citep{sheu2023a, magee2023a}.
Time-delay from multiply imaged supernovae  
can therefore be an important independent way to constrain \ho and address this tension in its measurements \citep[e.g.,][]{goldstein2017a, goldstein2018a, goldstein2019a, wojtak2019a, pierel2019a, suyu2020a}. In fact, \ed{\citet{kelly2023a} and \citet{pascale2024a} reported the first two measurements of \ho using lensed SNe.}
Moreover, time-delay \ho measurements are a powerful complement to other independent measurements of the dark energy equation of state
\citep[e.g.,][]{linder2011a, treu2016a, pierel2021a}.
\ed{Both for cluster/group and galaxy-scale lenses, to systematically discover lensed SNe, we can monitor a sample of strong lenses with the highest estimated supernova rates for the lensed sources \citep[e.g.,][]{shu2018a,shu2021a,suyu2023a}}.

\ed{Static strong lenses can be used to constrain cosmological parameters as well.
\citet{li2024a} showed a large sample, $\mathcal{O}(10^4)$,
in combination with velocity dispersion measurements, can provide competitive constraints, including for the possible evolution of dark energy \citep[e.g.,][]{chevallier2001a, linder2003a}.
Compound lenses, lensing systems with two or more sources at different redshifts, are extremely valuable. 
Even a relatively small sample constitutes a powerful cosmological probe \citep[e.g.,][]{collett2012a, collett2014a, linder2016a, sharma2022a, sharma2023a}. 
Compared with a typical strong lens, with a single source, these are much rarer, and yet there have been some very promising recent discoveries \citep{sheu2024a, dux2024a, bolamperti2024a}.}

Finally, for nearby strong lensing galaxies, 
extra-galactic tests of General Relativity can be performed 
by combining lens modeling with 
spatially resolved 
stellar kinematic observations \citep{collett2018a}. 

\rf{For many of these analyses, 
the available sample size of confirmed strong lenses is a major limiting factor.}
In the last few years, several groups have used convolutional neural networks to search for strong lensing systems in photometric surveys, including, in increasing sky coverage, 
CFHTLS \citep{jacobs2017a}, KiDS \citep{petrillo2017a, petrillo2019a, li2020a}, 
DES \citep{jacobs2019a, jacobs2019b}, and
Pan-STARRS \citep{canameras2020a}.

The DESI Legacy Surveys\footnote{\url{https://www.legacysurvey.org/}}
\citep{dey2019a},
for which at least $z$ band is observed with \ed{4-m telescopes},
covers $\sim$\coverage~deg$^2$,
almost four times the size of the Dark Energy Survey \citep{des2005a} footprint.
We identified over $\sim 3500$ new strong lenses in the Legacy Surveys \citep[][respectively; henceforth, H20, H21, S24]{huang2020a, huang2021a, storfer2024a} by using residual neural networks. 
We have also found 436 lensed quasars \citep{dawes2023a} using autocorrelation for a lensed quasar candidate sample.
The entire catalog of these lens candidates can be found on our project website.\footnote{\url{https://sites.google.com/usfca.edu/neuralens/}}

With so many lens discoveries,
a fast and robust
modeling pipeline is needed.
We therefore developed \gigal \citep{gu2022a}\ed{: 
a GPU-accelerated}, fully forward-modeling Bayesian pipeline that can speed up the lens modeling time by two orders of magnitude or more. 
Recently, we applied \gigal to \ed{an observed} lensing system, 
DESI-253.2534+26.8843 \citep{cikota2023a}, 
making it the first real lens to be modeled by GPUs.

Here, we present the \projname project. 
With thousands of lens candidates, we can choose the best ones for the purposes of 1) detecting low-mass DM halos and 2) measuring \ho\ --- using the best lensed quasar systems and through a future targeted search for live lensed supernovae among lensing systems that are the most likely to host them.
For high resolution imaging, 
our \HST SNAP program (ID: 15867, PI: Huang) 
\ed{has observed a subset of the best candidates.}
In this paper --- Paper~I of this series --- 
we describe our \hst program.  
We will also present \ed{the first model using GPUs for a strong lensing system with \hst data.}
Spectroscopic observations are also being carried out.
The first \ed{spectroscopic} results from the Dark Energy Spectroscopic Experiment \citep[DESI;][]{desi2016a, desi2016b} strong lens program  will be presented in Paper~II of this series (Huang, Inchausti et al., in prep).
For 
\ed{$\sim 30\%$ of the systems (preliminary estimate)}, the source redshifts \ed{are too high, such that} the typical emission features (e.g., [\ion{O}{2}]) are beyond the \ed{DESI wavelength} range.
For these, our near-IR spectroscopic results from the Keck~2 Telescope will be reported in Paper~III in this series (Agarwal et al., in prep).

This paper is organized as follows.  
We present our \HST SNAP program in   
\S\,\ref{sec:ls-hst}, followed by a brief description of the 
DESI and Keck spectroscopic observations in \S\,\ref{sec:spect}.
In \S\,\ref{sec:lens-model}, we present the model for one of our \hst systems using \gigal. \ed{Our findings are discussed in \S\,\ref{sec:discussion} and the conclusion is provided in \S\,\ref{sec:conclusion}.}

\FloatBarrier
\section{\HST SNAP Program}\label{sec:ls-hst}
In this section, we briefly describe our lens discoveries in the DESI Legacy Imaging Surveys in \S\,\ref{sec:discovery} and present the follow-up \hst SNAP program observations in \S\,\ref{sec:hst-obs}.
All the {\it HST} data used in this paper can be found in MAST: \dataset[10.17909/hx0v-9260]{http://dx.doi.org/10.17909/hx0v-9260}.
\subsection{Lens Discoveries in the DESI Legacy Imaging Surveys }\label{sec:discovery}

The DESI Legacy Imaging Surveys (hereafter Legacy Surveys) \ed{are} composed of three surveys: 
the Dark Energy Camera Legacy Survey (DECaLS), the Beijing-Arizona Sky Survey (BASS), and the Mayall $z$-band Legacy Survey (MzLS). 
DECaLS is observed by the Dark Energy Camera \citep[DECam;][]{flaugher2015a} on the 4-m Blanco telescope, which covers $\sim 9000$~deg$^2$ of the sky in the range of $-18^{\circ} \lesssim \delta \lesssim +32^{\circ}$. 
\ed{The MzLS has imaged the $\delta \gtrsim +32^\circ$ sub-region ($\sim5000$~deg$^2$) in $z$-band by the Mosaic3 camera \citep[][]{dey2016a} that complemented the BASS $g$- and $r$-band observations in the same \ed{sub-region} on 90Prime camera \citep[][]{williams2004a} on the Bok 2.3-m telescope.
}
\ed{Data Releases} 9 \& 10 (DR9, 10; Schlegel et al., in prep)
contain additional DECam data reprocessed from the Dark Energy Survey \citep[DES;][]{abbott2018a} for $\delta\lesssim-18^{\circ}$.
This provides an extra $\sim 5000~$deg$^2$, resulting in a total footprint of $\sim19,000$~deg$^2$. 
The Legacy Surveys \ed{are} imaged 
\ed{to a total depth of at least} 22.5 AB mag in $z$-band (for galaxies with an exponential disk profile \ed{and a half-light radius r$_{half} = 0.45\twopr$}). 
The \ed{average} FWHMs \ed{for} the delivered images are: $1.29\twopr$ ($g$), 1.18$\twopr$ ($r$), and 1.11$\twopr$ ($z$) for DECaLS; $1.61\twopr$ ($g$) and $1.47\twopr$ ($r$) for BASS; and $1.01\twopr$ ($z$) for MzLS.  
A more detailed description of the Legacy Surveys can be found on the Legacy Surveys website.\footnote{\href{https://www.legacysurvey.org/dr9/description/}{https://www.legacysurvey.org/dr9/description/}} 
On this dataset, 
we have performed three lens searches in the Legacy Surveys in \ed{Data Releases~7, 8, and~9} (H20, H21, S24, respectively).
We used a residual neural network architecture first developed by \citet{lanusse2018a}, and improved upon by H21 (the ``shielded" model).
For all three searches, we used observed images of both lenses and non-lenses (instead of simulations or a combination of observations and simulations) for training.
\subsection{\HST Observations}\label{sec:hst-obs}
We submitted a subset of our most promising lens candidates (112 targets) for the \emph{Hubble} SNAP program GO-15867 (PI: Huang),
\emph{Confirming Strong Galaxy Gravitational Lenses in the DESI Legacy Imaging Surveys}.
\ed{\hst not only provides higher angular resolution, but reaches fainter surface brightness due to the fainter background in space. 
This can reveal the presence of lensed galaxies with lower surface brightness, including those at higher redshifts that have been dimmed by $(1+z)^4$.
In addition, a NIR filter was chosen to make possible the detection of higher redshift sources that are beyond the optical ($grz$) reach of the DESI Legacy Surveys.
These two factors expand the depth
and redshift range of detectable arcs. 
Indeed, for a large number of systems in this program, the \hst NIR observations revealed additional lensed sources, often with a larger Einstein radius than the one seen in the respective discovery $grz$ images.}
The targets were chosen for two main science goals.
One, identifying the best systems for a future targeted search for lensed supernovae to measure the Hubble constant \citep[e.g.,][]{shu2018a, shu2021a}.
Lensed supernovae have been discovered in galaxy-scale \citep[e.g.,][]{goobar2023a, pierel2023a}, 
and cluster-scale lenses \citep[e.g.,][]{kelly2023a, frye2023a}.
Two, low-mass halo ($<10^9 M_\odot$) detection.
This is a powerful test of the cold dark matter (CDM) model, which predicts an abundance of low-mass halos, as subhalos or line-of-sight interlopers.
These do not host a galaxy and hence are invisible.
Beyond our local universe, they  are only detectable by their gravitational effect, especially in strong lensing systems \citep[e.g.,][]{vegetti2010a, cagan-sengul2022a}.
For this purpose, systems with a single galaxy as the main lens is usually preferred,
as these are considered to be easier to model and hence can yield low-mass halo detection with greater confidence.
\ed{However}, cluster-scale lenses have been used for this purpose too \citep[e.g.,][]{dai2020a}. 
To the degree possible, using the ground-based Legacy Surveys images, we chose half of our \hst targets to be galaxy-scale lenses and  
the other half, group/small cluster lenses.
Among strong lenses, being in-between galaxy-scale and cluster-scale, the latter category is somewhat understudied.
\ed{For example, in the SL2S program, 26 group-scale lenses were followed up with \hst observations \citep[][Table~3]{more2012a}.
Our program in the end observed a comparable number of group/small cluster lenses.}
This is likely due to the fact that these systems are outside of the previous selection windows: galaxy-scale lenses were discoverable in highly multiplexed fiber spectroscopy and cluster-scale lenses are more readily identified \ed{by eye in deep images}.
\ed{In contrast}, neural net based search methods applied to ground-based observations have discovered an abundance of strong lenses at the group to small cluster scales.
We would like to explore its potential for these two science goals.

All targets have $3 \times$399.23~sec exposure, 
for a total of 1197.7~sec, 
on WFC3 using the F140W filter in the NIR channel.
The targets were approximately centered in the WFC3 aperture, with no CR split, because we wanted to keep read noise down. 
Of the 112 systems submitted to \hst, 
a total of 53 were targeted. 
Two of them were not successfully observed due to the loss of the guide star.
Of the 51 systems successfully observed, \hst confirmed all of them to be strong lenses, 
by revealing one or more arcs in each.\footnote{\ed{Of these}, DESI-118.8480+34.7610 is a known system \citep[][SDSS~J0755+3445]{shu2016b}, 
and it was observed in the optical by \hst WFC3 using UVIS F606W (GO-14189, PI: A. Bolton).}
The native pixel size is $0.13\twopr$, and each image is drizzled to $0.065''$.
In this \emph{Hubble} sample, 
\href{https://www.legacysurvey.org/viewer/?ra=025.4848&dec=+30.7585&layer=ls-dr9&pixscale=0.262&zoom=16}{DESI-025.4848+30.7585}\footnote{For the first mention of a system, we hyperlink the name to its Legacy Surveys DR9 image.}  has the faintest arcs.
It has a lens redshift of $z_d = 0.524$ from the extended Baryon Oscillation Spectroscopic Survey (eBOSS), 
which is further confirmed by DESI spectroscopy.
In the Legacy Surveys discovery image (see Figure~\ref{fig:hstpanel-1}), 
there appear to be a lensed arc (arc~1) to the SE of the elliptical galaxy at the center and a hint of a small arc (arc~2) to the NW of and just above the same elliptical galaxy.
Both are confirmed in the \hst image.
But \hst also reveals that just below arc~2, even closer to the lens, there appears to be a third arc (arc~3).
DESI provides the redshift for arc~1 ({$z_{s, 1} = 1.2223$, based on clear detection of the [\ion{O}{2}] doublet emission from DESI Year 1 data).
\citet{talbot2021a} analyzed the lens spectrum closely and found evidence for emission lines of a lensed source in the same eBOSS fiber, 
and determined its redshift at $z_{s, 2} = 1.216$.
The eBOSS fiber is $2''$ in diameter \citep{dawson2016a},
and thus is capable of detecting emission features for an arc $\lesssim 1''$ away from the lens.
Therefore this redshift corresponds to either arc~3 ($0.87''$ from the lens) or arc~2 (at a distance of $1.20''$).
Arc~1 at 2.10$''$ is not likely.
In addition to this system, we will also provide spectroscopic redshifts from DESI and Keck for  DESI-165.4754-06.0423 (see Figure~\ref{fig:hstpanel-2}), 
for which we will show a lens model in \S\,\ref{sec:lens-model}.
DESI redshifts for other systems will be presented in Paper~II and future publications.

Figure~\ref{fig:hstpanel-1} shows the first eight systems in this sample.
These eight are representative of the type of lensing configuration for this \hst sample.
We show the rest of the 51 systems in Figures~\ref{fig:hstpanel-2}~-~\ref{fig:hstpanel-6}. 
There are a number of striking systems.
In Figure~\ref{fig:hstpanel-1}, 
\ed{for \href{https://www.legacysurvey.org/viewer/?ra=4.2564&dec=-10.1530&layer=ls-dr9&pixscale=0.262&zoom=16}{DESI-004.2564-10.1530} and \href{https://www.legacysurvey.org/viewer/?ra=23.6765&dec=+4.5639&layer=ls-dr9&pixscale=0.262&zoom=16}{DESI-023.6765+04.5639}}, 
both group lenses, 
compared with the discovery images from the Legacy Surveys in the optical bands of $grz$,
the \hst NIR images reveal additional fainter arcs in each. 
This is true for a number of group-scale lensing systems in this sample.
\href{https://www.legacysurvey.org/viewer/?ra=72.0873&dec=-19.4174&layer=ls-dr9&pixscale=0.262&zoom=16}{DESI-072.0873-19.4172} (Figure~\ref{fig:hstpanel-2}) is an Einstein cross with a remarkable average lens-image separation of 3.95$\twopr$.
The photometric redshift for the lens is $z_{d, phot} = 0.967\pm0.167$.\footnote{Here and later in this paper, 
the photometric redshifts are from \citet{zhou2020a}.} 
There appear to be at least three other elliptical galaxies with similar colors (and therefore similar redshifts) within or near the Einstein cross, 
though they are considerably smaller. 
Inspection of the environment around this system shows that the lensing galaxy is at the center of a high redshift cluster.
\ed{The presence of the Einstein cross, rarely seen for a group/cluster scale lens \citep[e.g.,][which shows an Einstein cross for a cluster lens with $z_d = 0.49$]{sheu2024a}, indicates that mass distribution is relaxed and highly symmetric within the Einstein radius, perhaps all the more unusual at such a high redshift. The modeling of this system is currently underway.}
Other notable systems include: 
a galaxy-scale lensing system that possibly has two multiply-imaged sources (\href{https://www.legacysurvey.org/viewer/?ra=133.3800&dec=+23.3653&layer=ls-dr9&pixscale=0.262&zoom=16}{DESI-133.3800+23.3652} in Figure~\ref{fig:hstpanel-2}, \ed{shown with arrows})
and two group-scale systems with multiple images of a background galaxy with complex structure 
(\href{https://www.legacysurvey.org/viewer/?ra=189.5370&dec=15.0309&layer=ls-dr9&pixscale=0.262&zoom=16}{DESI-189.5370+15.0309} in Figure~\ref{fig:hstpanel-3} with a doubly imaged background spiral galaxy, 
or two half ``cinnamon buns", and 
\href{https://www.legacysurvey.org/viewer/?ra=327.8408&dec=13.7884&layer=ls-dr9&pixscale=0.262&zoom=16}{DESI-327.8408+13.7884} in Figure~\ref{fig:hstpanel-6}, showing great details of a triply lensed spiral galaxy, \ed{which suggests a possible naked cusp configuration \citep[e.g.,][]{lewis2002a}}.
Finally, the \hst image of \href{https://www.legacysurvey.org/viewer/?ra=254.4235&dec=34.8161&layer=ls-dr9&pixscale=0.262&zoom=16}{DESI-254.4235+34.8162} (Figure~\ref{fig:hstpanel-5}) possibly reveals a radial counter arc very close to the lensing galaxy.

\clearpage




\begin{minipage}{\linewidth}
\makebox[\linewidth]{  \includegraphics[keepaspectratio=true,scale=0.45]{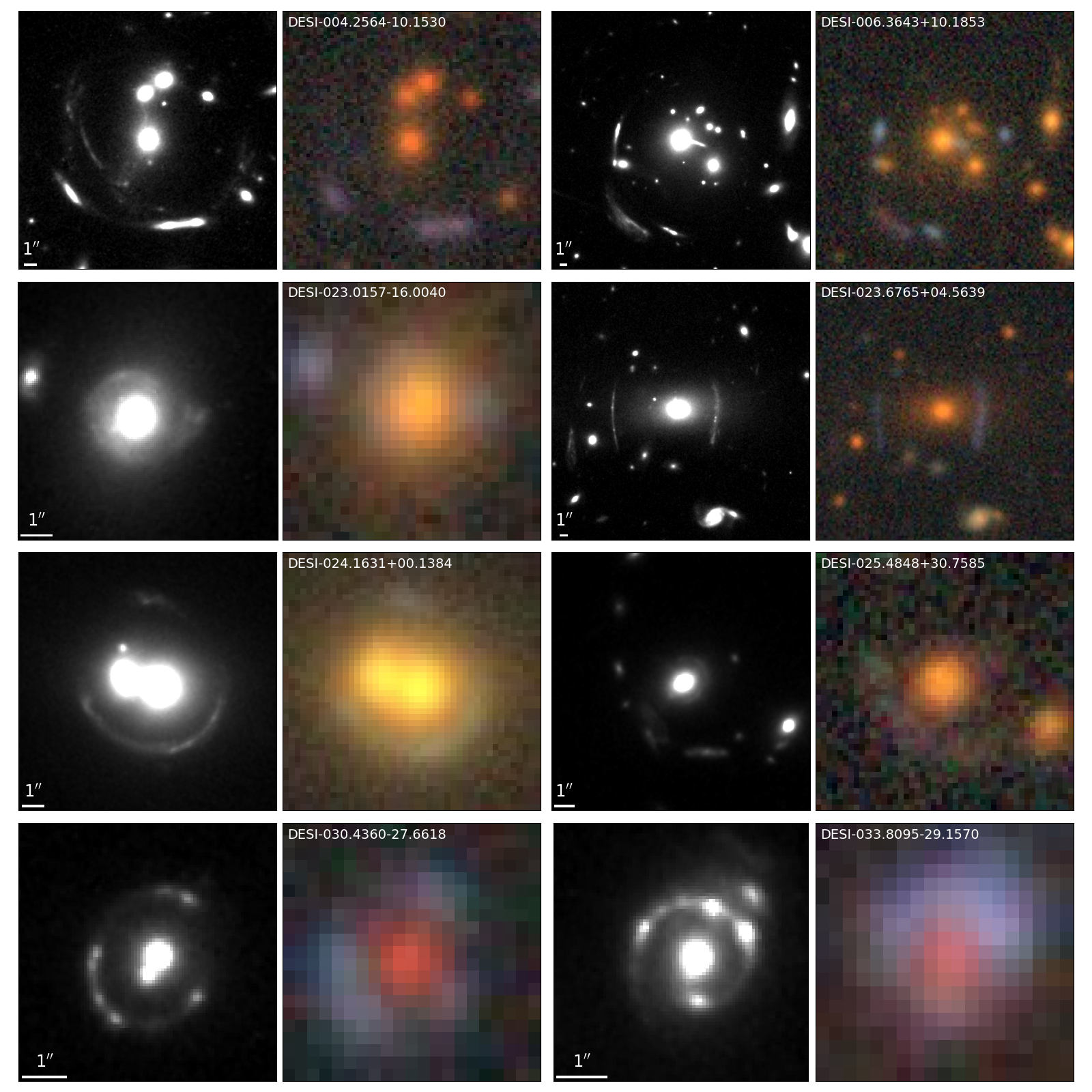}}
\captionof{figure}{
The first eight of the 51 systems observed by the \hst SNAP program GO-15867. 
All were confirmed to be strong lenses by \hst with arc(s) in each clearly revealed.
For this and the next five figures (up to and including Figure~\ref{fig:hstpanel-6}): 
The naming convention is RA and Dec in decimal format.
North is up, and East to the left. 
The systems are arranged in ascending RA.
For each, we show the \hst WFC3 NIR F140W image (left)
and the Legacy Surveys optical image in $grz$ bands (right; these were the discovery images).
The eight systems in this panel are representative of the 51 systems in this \hst program: There are three group/small cluster scale lenses, DESI-004.2564-10.1530, DESI-006.3643+10.1853, and DESI-023.6765+04.5639.
For each of them, there are at least two sets of arcs. 
Compared with the Legacy Surveys optical image, 
the NIR \hst image revealed new arcs for the first and third system.
\ed{DESI-023.0157-16.0040, DESI-025.4848+30.7585, and DESI-033.8095-29.1570 are galaxy-scale lens} (for more on the lensing nature of DESI-025.4848+30.7585, see text).
Finally, DESI-024.16341+00.1384 and DESI-030.4360-27.6618 \ed{each have} two very close-by elliptical galaxies as the lens.}
\label{fig:hstpanel-1}
\end{minipage}

\begin{minipage}{\linewidth}
\makebox[\linewidth]{  \includegraphics[keepaspectratio=true,scale=0.45]{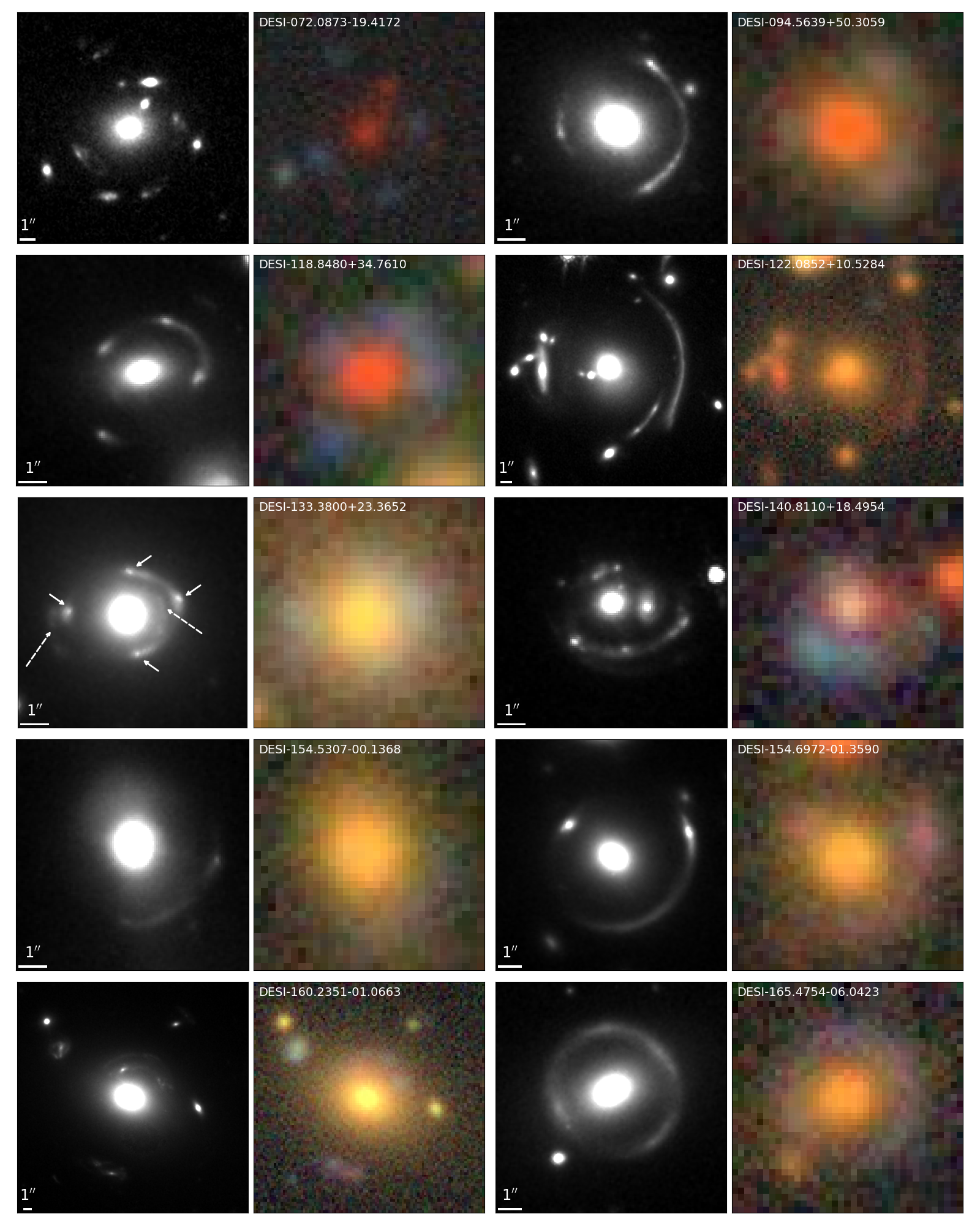}}
\captionof{figure}{
The next 10 (\pound9 -- \pound18) of the 51 systems in \hst GO-15867. 
For details, see the caption of Figure~\ref{fig:hstpanel-1}. 
The lens model for DESI-165.4754-06.0423 is presented in \S\,\ref{sec:lens-model}.
\ed{
DESI-133.3800+23.3652 is a galaxy-scale lens, likely with two sources: one quadruply lensed (solid arrows) and one doubly lensed (dashed arrows).
}
}
\label{fig:hstpanel-2}
\end{minipage}

\begin{minipage}{\linewidth}
\makebox[\linewidth]{  \includegraphics[keepaspectratio=true,scale=0.45]{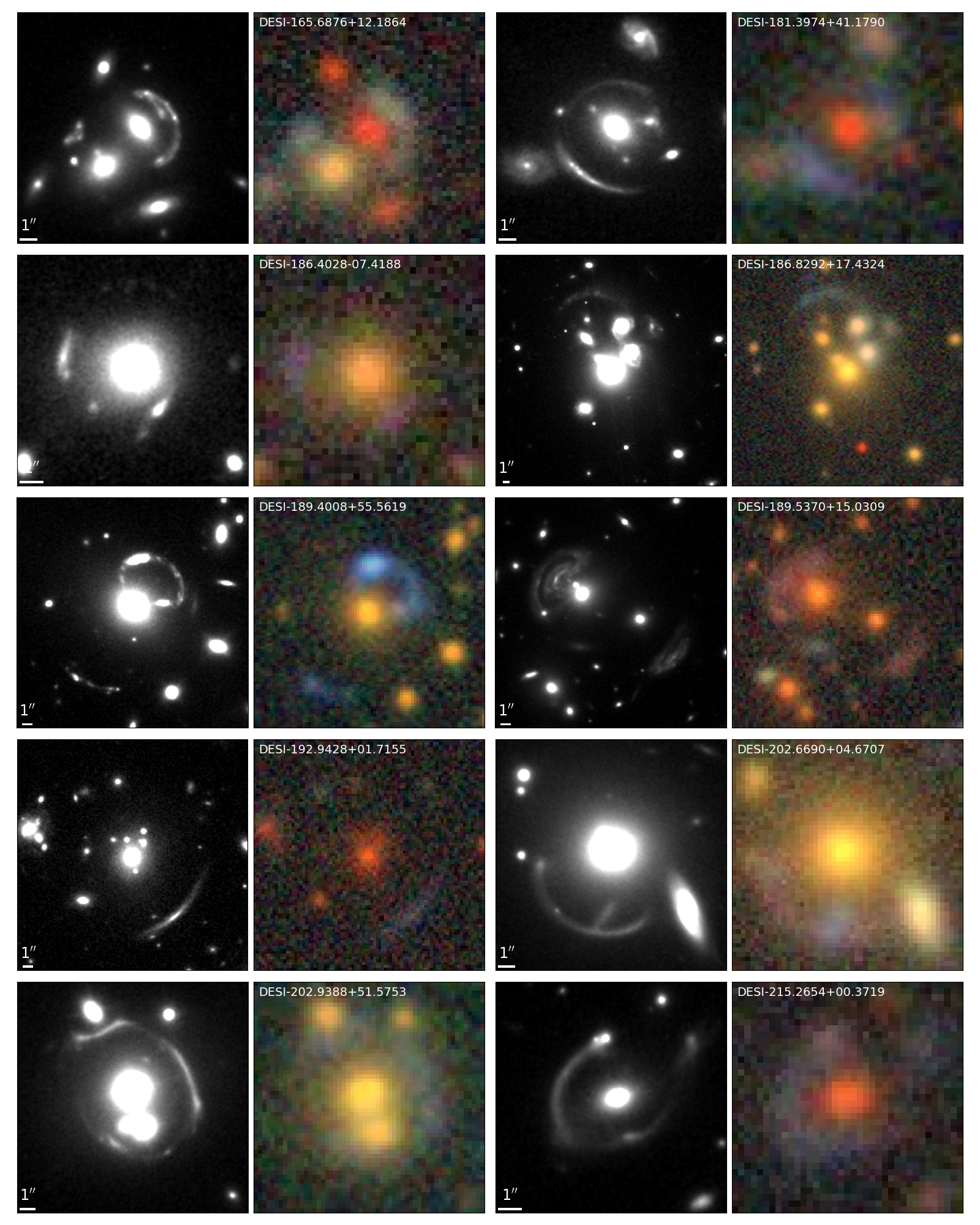}}
\captionof{figure}{
Systems \pound19 - \pound28 of the 51 in \hst GO-15867. 
For details, see the caption of Figure~\ref{fig:hstpanel-1}.
Note that DESI 189.5370+15.0309 shows a doubly imaged background spiral galaxy with complex structure, 
resembling two half ``cinnamon buns", 
with different sizes.}
\label{fig:hstpanel-3}
\end{minipage}

\begin{minipage}{\linewidth}
\makebox[\linewidth]{  \includegraphics[keepaspectratio=true,scale=0.45]{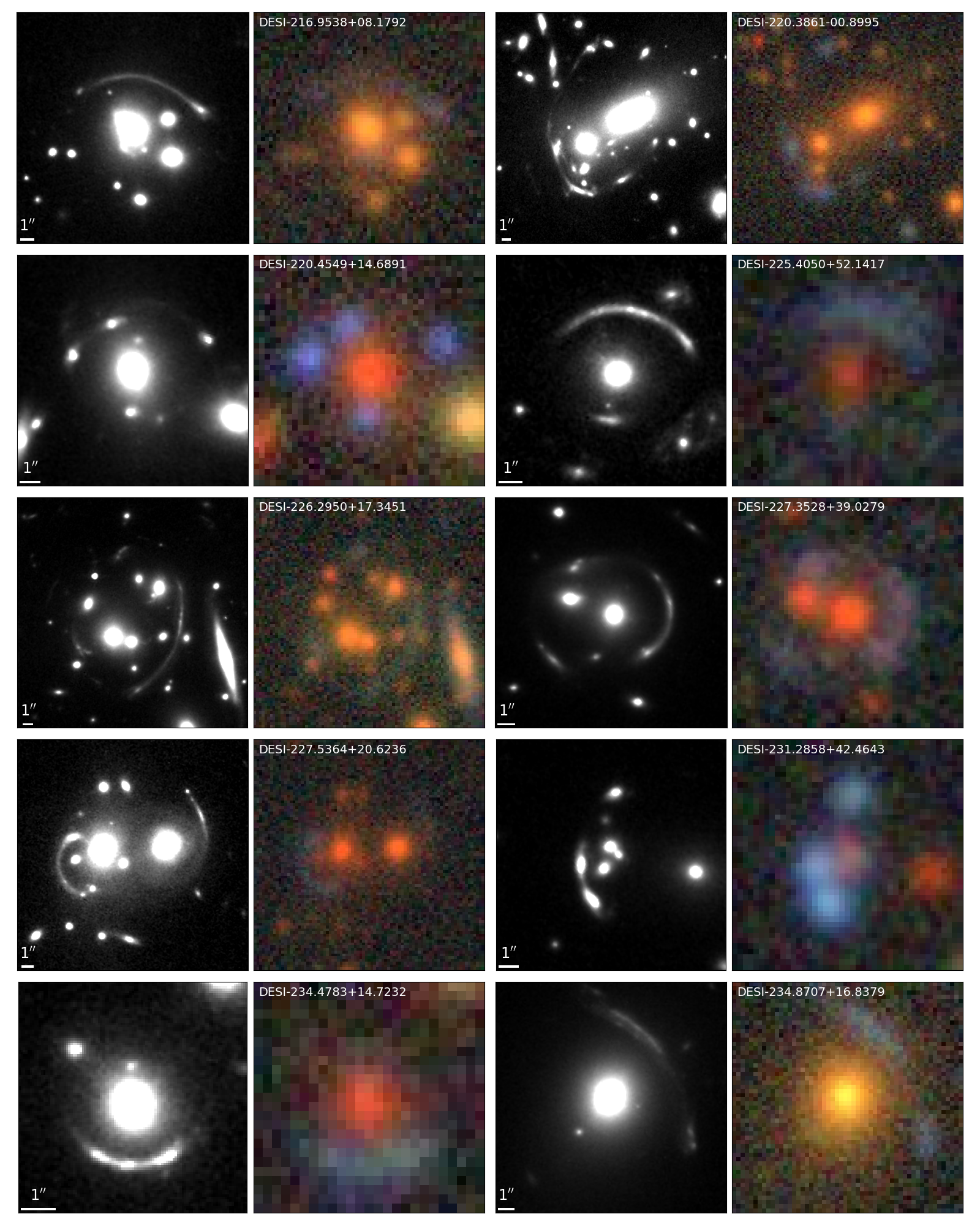}}
\captionof{figure}{
Systems \pound29 -- \pound38 of the 51 in \hst GO-15867. For details, see the caption of Figure~\ref{fig:hstpanel-1}.}
\label{fig:hstpanel-4}
\end{minipage}

\begin{minipage}{\linewidth}
\makebox[\linewidth]{  \includegraphics[keepaspectratio=true,scale=0.45]{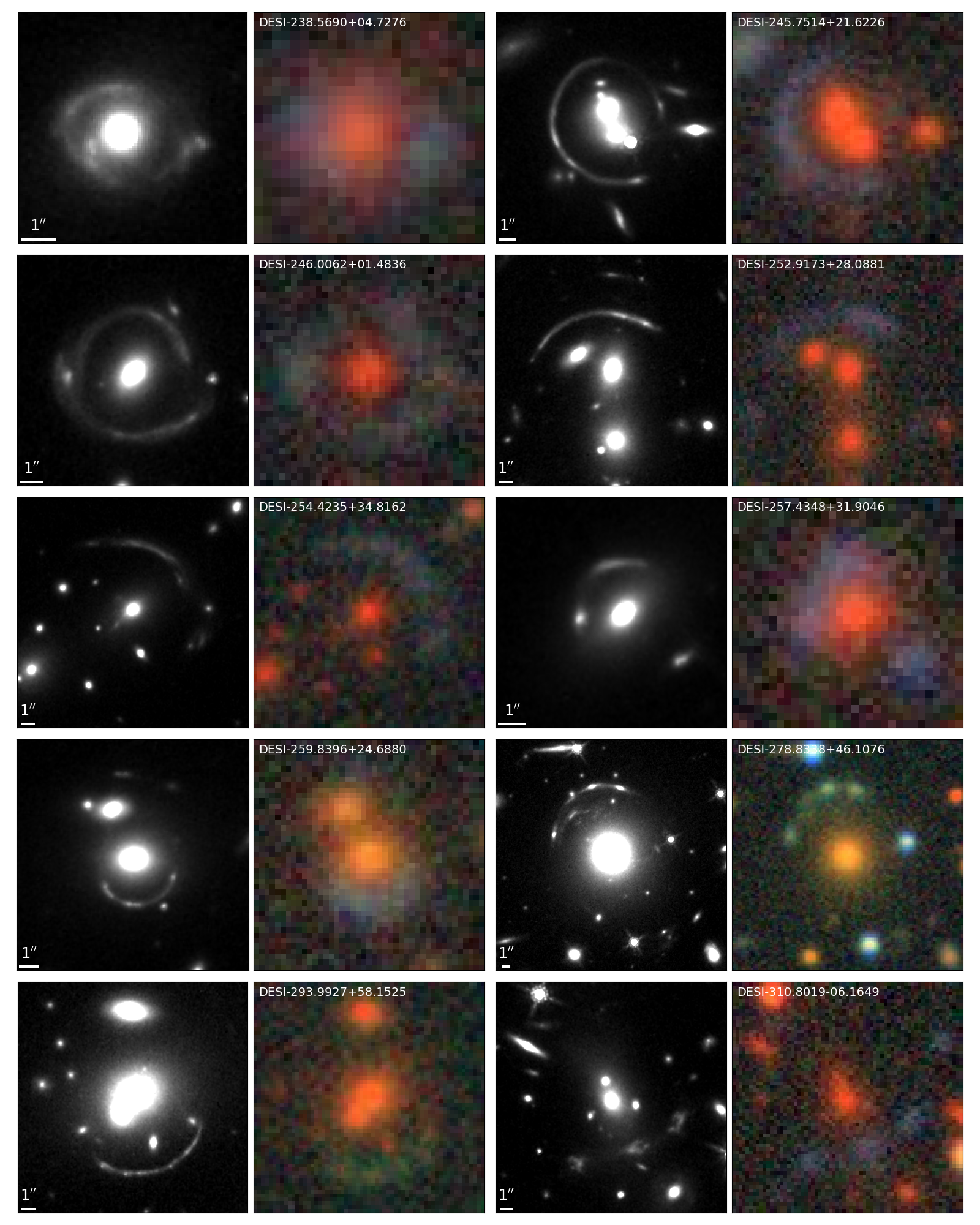}}
\captionof{figure}{
Systems \pound39 -- \pound48 of the 51 in \hst GO-15867. 
For details, see the caption of Figure~\ref{fig:hstpanel-1}.
Note DESI-238.5690+04.7276, a galaxy-scale lens, 
likely with two lensed sources,
which are quadruply (fainter) and doubly (brighter) lensed, respectively.
}
\label{fig:hstpanel-5}
\end{minipage}

\begin{minipage}{\linewidth}
\makebox[\linewidth]{  \includegraphics[keepaspectratio=true,scale=0.45]{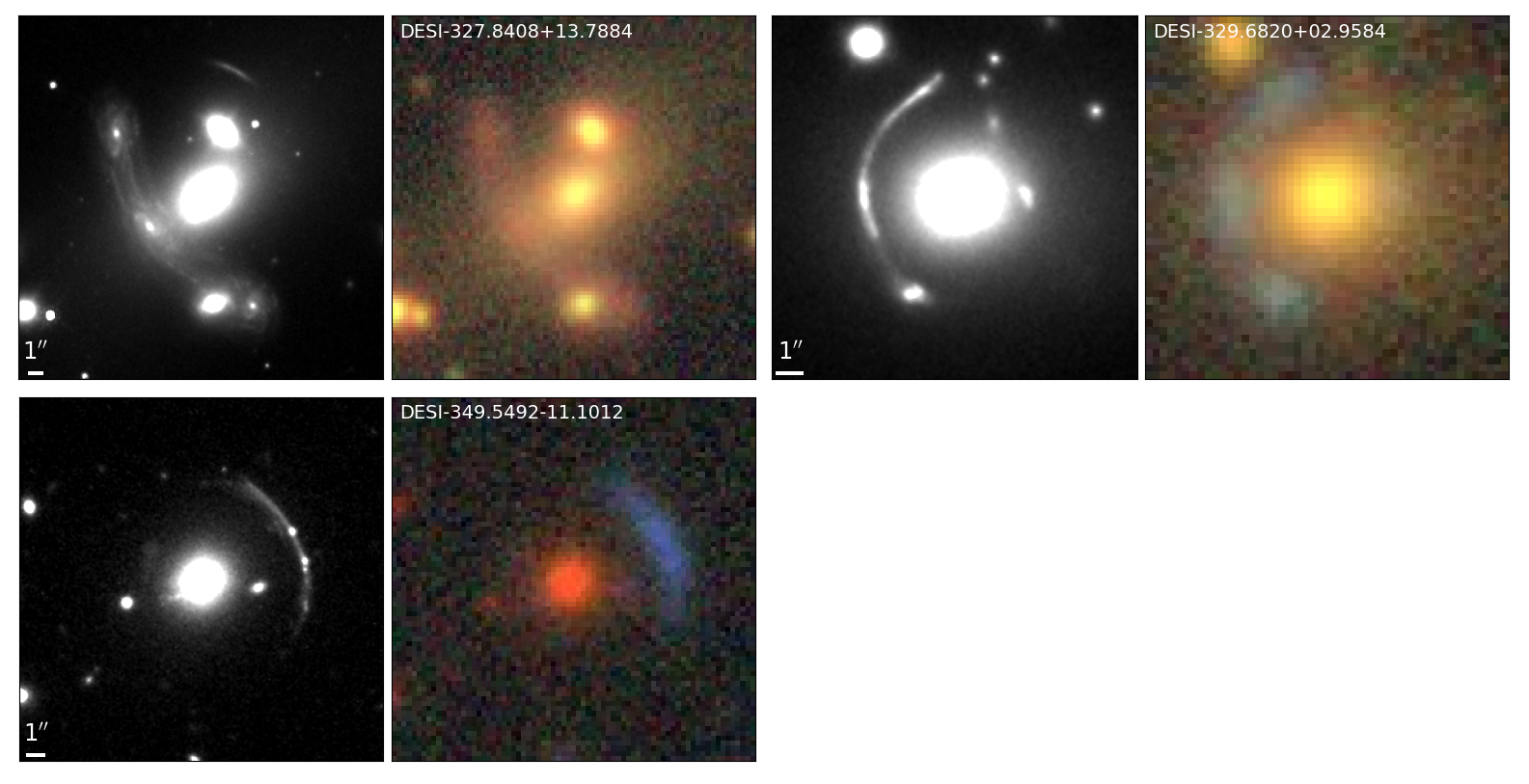}}
\captionof{figure}{
Systems \pound49 - \pound51 of the 51 in \hst GO-15867. 
For details, see the caption of Figure~\ref{fig:hstpanel-1}.
Note that DESI-327.8408+13.7884 shows a triply lensed galaxy around a group lens \ed{(see text)}, revealing great details of the magnified background spiral galaxy.}
\label{fig:hstpanel-6}
\end{minipage}


\clearpage

We present \hst photometry for the brightest lensed source images \ed{for the source identified in the discovery image}\footnote{\ed{As noted above, in some cases, \hst NIR observations revealed additional lensed sources. These are typically significantly fainter and their photometry is not presented here.}} for all 51 systems in Table~\ref{tab:hst-phot}, 
\ed{where we define the contour to be
at 75\% of the peak pixel value}.
To account for the luminosity contribution from the lens we employ GALFIT \citep{Peng2010a}, a parametric 2D fitting algorithm.  
We model the 
light components of the lensing galaxy using 
the S\'ersic luminosity profile \citep{sersic1963a}. Each S\'ersic component in the fitting algorithm is described by the \ed{7 parameters}: $(x,y)$ position, integrated magnitude $m$, effective radius $r_e$, concentration index $n$, axis ratio $q$, and position angle $\theta$.
\par Prior to the fitting procedure, a PSF was generated for each image by stacking star cutouts from the field and iteratively refining a proposed PSF model based on the stack. This process was carried out using \texttt{PSFr} 
in the \texttt{Lenstronomy} package \citep{birrer2021a}. Following successful extraction of PSFs, each galaxy was initially fit \ed{using} a single S\'ersic component, with additional components added when necessary to adequately subtract the lens light from the image. Goodness of fit was determined using the $\chi^2$ metric as well as visual inspection of the image residuals. In cases where model parameters deviated from physically acceptable values, such as extreme $r_e$, constraints were imposed. For systems where the lensing field was densely populated, the light profiles of all galaxies in the image region were also modeled and subtracted to reduce contamination of source light. Moreover, in certain outlier cases where 
two source images appeared in close proximity to each other, the source light for such systems was also modeled in order to properly isolate each source image. Source light distribution was modeled using \ser profiles along with an additional bending mode where necessary.

\startlongtable
\begin{deluxetable*}{lcccc}
\tabletypesize{\scriptsize}
\tablecaption{Isophotal magnitudes for the brightest source image
\label{tab:hst-phot}}
\tablehead{
    \colhead{Target Name} &
    \colhead{\ed{RA (deg)}} &
    \colhead{\ed{Dec (deg)}} &
    \colhead{F140W Isophotal Magnitude \ed{(AB mag)}} &
    \colhead{Contour Area \ed{(arcsec$^2$)}} 
}
\startdata
DESI-004.2564-10.1530 & 4.2560 & $-$10.1549 & 23.51 $\pm$ 0.02 & 0.0380 \\
DESI-006.3643+10.1853 & 6.3646 & 10.1820 & 23.79 $\pm$ 0.02 & 0.0338 \\
DESI-023.0157-16.0040 & 23.0152 & $-$16.0041 & 25.04 $\pm$ 0.04 & 0.0634 \\
DESI-023.6765+04.5639 & 23.6753 & 4.5635 & 25.68 $\pm$ 0.05 & 0.0338 \\
DESI-024.1631+00.1384 & 24.1628 & 0.1376 & 24.89 $\pm$ 0.04 & 0.0380 \\
DESI-025.4848+30.7585 & 25.4846 & 30.7587 & 26.43 $\pm$ 0.07 & 0.0423 \\
DESI-030.4360-27.6618 & 30.4365 & $-$27.6617 & 25.43 $\pm$ 0.05 & 0.0211 \\
DESI-033.8095-29.1570 & 33.8094 & $-$29.1571 & 24.16 $\pm$ 0.03 & 0.0169 \\
DESI-072.0873-19.4172 & 72.0883 & $-$19.4177 & 26.20 $\pm$ 0.06 & 0.0296 \\
DESI-094.5639+50.3059 & 94.5634 & 50.3065 & 24.88 $\pm$ 0.04 & 0.0211 \\
DESI-118.8480+34.7610 & 118.8477 & 34.7615 & 24.64 $\pm$ 0.03 & 0.0338 \\
DESI-122.0852+10.5284 & 122.0836 & 10.5279 & 22.43 $\pm$ 0.01 & 0.5197 \\
DESI-133.3800+23.3652 & 133.3795 & 23.3653 & 23.98 $\pm$ 0.02 & 0.0296 \\
DESI-140.8110+18.4954 & 140.8114 & 18.4951 & 24.35 $\pm$ 0.03 & 0.0423 \\
DESI-154.5307-00.1368 & 154.5299 & $-$0.1369 & 25.42 $\pm$ 0.05 & 0.0296 \\
DESI-154.6972-01.3590 & 154.6977 & $-$1.3587 & 23.24 $\pm$ 0.02 & 0.0296 \\
DESI-160.2351-01.0663 & 160.2353 & $-$1.0654 & 23.55 $\pm$ 0.02 & 0.0423 \\
DESI-165.4754-06.0423 & 165.4760 & $-$6.0425 & 23.87 $\pm$ 0.02 & 0.1394 \\
DESI-165.6876+12.1864 & 165.6870 & 12.1865 & 25.02 $\pm$ 0.04 & 0.0211 \\
DESI-181.3974+41.1790 & 181.3981 & 41.1783 & 24.47 $\pm$ 0.03 & 0.0465 \\
DESI-186.4028-07.4188 & 186.4026 & $-$7.4193 & 24.92 $\pm$ 0.04 & 0.0296 \\
DESI-186.8292+17.4324 & 186.8287 & 17.4343 & 24.13 $\pm$ 0.02 & 0.2113 \\
DESI-189.4008+55.5619 & 189.4036 & 55.5601 & 24.26 $\pm$ 0.03 & 0.0380 \\
DESI-189.5370+15.0309 & 189.5388 & 15.0313 & 23.81 $\pm$ 0.02 & 0.0676 \\
DESI-192.9428+01.7155 & 192.9417 & 1.7138 & 23.59 $\pm$ 0.02 & 0.1648 \\
DESI-202.6690+04.6707 & 202.6700 & 4.6698 & 23.74 $\pm$ 0.02 & 0.1310 \\
DESI-202.9388+51.5753 & 202.9371 & 51.5752 & 23.32 $\pm$ 0.02 & 0.1817 \\
DESI-215.2654+00.3719 & 215.2652 & 0.3715 & 24.39 $\pm$ 0.03 & 0.0845 \\
DESI-216.9538+08.1792 & 216.9525 & 8.1795 & 24.21 $\pm$ 0.03 & 0.0507 \\
DESI-220.3861-00.8995 & 220.3884 & $-$0.9006 & 23.94 $\pm$ 0.02 & 0.0507 \\
DESI-220.4549+14.6891 & 220.4557 & 14.6893 & 23.74 $\pm$ 0.02 & 0.0338 \\
DESI-225.4050+52.1417 & 225.4047 & 52.1424 & 24.97 $\pm$ 0.04 & 0.0380 \\
DESI-226.2950+17.3451 & 226.2932 & 17.3455 & 24.68 $\pm$ 0.03 & 0.0549 \\
DESI-227.3528+39.0279 & 227.3521 & 39.0273 & 23.91 $\pm$ 0.02 & 0.0507 \\
DESI-227.5364+20.6236 & 227.5339 & 20.6244 & 23.74 $\pm$ 0.02 & 0.1099 \\
DESI-231.2858+42.4643 & 231.2877 & 42.4639 & 23.08 $\pm$ 0.02 & 0.0465 \\
DESI-234.4783+14.7232 & 234.4780 & 14.7227 & 25.12 $\pm$ 0.04 & 0.0254 \\
DESI-234.8707+16.8379 & 234.8703 & 16.8391 & 24.47 $\pm$ 0.03 & 0.0803 \\
DESI-238.5690+04.7276 & 238.5695 & 4.7277 & 24.24 $\pm$ 0.03 & 0.1352 \\
DESI-245.7514+21.6226 & 245.7524 & 21.6224 & 23.71 $\pm$ 0.02 & 0.0887 \\
DESI-246.0062+01.4836 & 246.0060 & 1.4843 & 24.97 $\pm$ 0.04 & 0.0803 \\
DESI-252.9173+28.0881 & 252.9166 & 28.0890 & 24.45 $\pm$ 0.03 & 0.0634 \\
DESI-254.4235+34.8162 & 254.4234 & 34.8174 & 24.00 $\pm$ 0.02 & 0.1732 \\
DESI-257.4348+31.9046 & 257.4352 & 31.9046 & 24.13 $\pm$ 0.02 & 0.0380 \\
DESI-259.8396+24.6880 & 259.8397 & 24.6873 & 24.94 $\pm$ 0.04 & 0.0338 \\
DESI-278.8338+46.1076 & 278.8348 & 46.1099 & 23.52 $\pm$ 0.02 & 0.0423 \\
DESI-293.9927+58.1525 & 293.9930 & 58.1510 & 26.15 $\pm$ 0.06 & 0.0211 \\
DESI-310.8019-06.1649 & 310.8015 & $-$6.1660 & 23.86 $\pm$ 0.02 & 0.1648 \\
DESI-327.8408+13.7884 & 327.8425 & 13.7896 & 23.67 $\pm$ 0.02 & 0.0465 \\
DESI-329.6820+02.9584 & 329.6824 & 2.9595 & 23.91 $\pm$ 0.02 & 0.0592 \\
DESI-349.5492-11.1012 & 349.5478 & $-$11.1005 & 25.33 $\pm$ 0.04 & 0.0085 \\
\enddata
\end{deluxetable*}

\section{Spectroscopic Observations}\label{sec:spect}
The DESI Strong Lens Secondary Target program \citep[among a number of ``secondary target'' programs, see][]{desi2023a} has \ed{so far} observed a \ed{majority} of our lens candidates (Paper~II).
The success rate of obtaining redshifts for lensing galaxies by DESI is \ed{very} high \ed{($> 90\%$)\footnote{\ed{The reasons for redshift failures are: low signal-to-noise ratio, 
contamination by the presence of source spectral features (typically strong emission lines), spectral reduction issues, and Redrock (the spectral fitting pipeline for DESI) fitting failure.}}}.
These are nearly always bright elliptical galaxies with a number of identifiable features in the optical range,
typically absorption lines
and the 4000 \ang break. 
For lensed sources, 
some of the spectra have no clear features,  
likely due to the fact that they are too faint and/or their redshifts place key spectral features outside the optical range. 
Lensed sources are typically star forming galaxies for which the [\ion{O}{2}] doublet emission feature is often used to anchor redshift fits. 
This makes redshift measurements in the optical range for $z_s \gtrsim 1.6$ challenging.\footnote{It is true that for $z_s \gtrsim 2.0$, 
Ly-$\alpha$ will 
be in the optical range, 
though
this feature is not always present.}
Thus for some systems obtaining $z_s$
may require NIR spectroscopic follow-up observations. 
Our on-going Keck NIRES program has determined the source redshifts for six of them (Paper~III),
for which $z_s \gtrsim 1.6$.
One of these systems is DESI-165.4754-06.0423. 
Below we present the lens model for this system.

\section{Lens Modeling}\label{sec:lens-model}

As a demonstration, we use \gigal \citep{gu2022a} to model one lens system observed by our DESI and Keck NIRES programs, \href{https://www.legacysurvey.org/viewer/?ra=165.4753&dec=-6.0424&layer=ls-dr9&pixscale=0.262&zoom=16}{DESI-165.4754-06.0423}.
The modeling of other systems with \hst data will be presented in future publications.
DESI-165.4754-06.0423 was discovered in H21, with a numerical grade of 2.5 (out of 4), corresponding to a C~grade 
(In H21, we mentioned that all candidates with a human inspection score $\geq 2.5$ are likely lensing systems).
The lens and source redshifts of $z_d = 0.4834$ and
$z_s = 1.6748$ are obtained from DESI and Keck NIRES, respectively (Paper II \& III).
Based on color and photo-$z$, 
this galaxy appears to be at the center, 
and the brightest member (18.8~mag in $r$-band), of a small group.
While most other group members are 12$\twopr$ or more away, 
there is a small nearby galaxy with $z_{phot} = 0.579\pm 0.166$, consistent with it being a group member, 
with a $r$-band magnitude of 23.2~mag.
We will refer to this object as the ``nearby galaxy''.
The Legacy Surveys image of DESI-165.4754-06.0423 shows an elliptical galaxy surrounded by what appears to be an almost complete Einstein ring, with an Einstein radius of approximately $\sim 2.5''$.
The \hst image showed that it is a quadruply lensed system.

\gigal is a fully forward-modeling Bayesian lens modeling pipeline.
Briefly, it consists of three steps: finding the maximum a posteriori (MAP) for the lensing parameters via multi-start gradient descent, determining a surrogate multidimensional Gaussian covariance matrix for these parameters using stochastic variational inference (SVI), and finally sampling with Hamiltonian Monte Carlo (HMC).
All three steps use gradient descent with automatic differentiation and take advantage of GPU acceleration.  
Unlike with other lens modeling pipelines, 
we do not take a staged approach, 
which may first fit for and then subtract the lens light, 
followed by initializing with a simple model and gradually increasing the model complexity, typically fixing the values of some parameters in this process (e.g., \ser index values).
Instead, to avoid landing in a local minimum and to account for full statistical uncertainties, we perform full forward modeling.
It is robust and very fast (\ed{$\sim 3$~hours even for a challenging system such as this one, as we will show below).}


We show this system in Figure~\ref{fig:what-to-mask} again, 
although with a different orientation from Figure~\ref{fig:hstpanel-2}.
We use a $27 \times 27$ pixel empirical PSF generated by stacking PSF-like objects identified by SExtractor.\footnote{We use the implementation provided in \url{https://github.com/sibirrer/AstroObjectAnalyser/}.}
Our mass model comprises an elliptical power law (EPL) for the main lens with external shear.
\ed{The EPL model is characterized by the surface mass density expressed in units of the critical density, commonly referred to as convergence,
\begin{equation}
    \kappa(x_{lens}, y_{lens}) = \frac{3-\gamma}{2}\left(\frac{\theta_E}{\sqrt{q x_{lens}^2+y_{lens}^2 / q}}\right)^{\gamma-1},
\end{equation}
where $\theta_E$ is the Einstein radius (in the intermediate-axis convention), $\gamma$ is the deprojected 3D mass profile slope \citep[e.g.,][]{tessore2015a}, $(x_{lens}, y_{lens})$ are lens-centric coordinates, and $q$ is the axial ratio.
In lens modeling, often $q$ and the position angle $\phi$ are reparametrized as eccentricities,
\begin{equation}
    (\epsilon_1, \epsilon_2) = \frac{1-q}{1+q}\left(cos(2\phi), sin(2\phi)\right).
\end{equation}
The external shear is characterized by $\gamma_{ext, 1}$ and $\gamma_{ext, 2}$.}



We model the lens light with two elliptical \ser profiles\ed{,
defined as,
\begin{equation}
I(x_{light}, y_{light})= I_0 \ exp\left(-b_n\left(\left(\frac{\sqrt{q x_{light}^2+y_{light}^2 / q}}{R}\right)^{1/n} - 1
\right)
\right),
\end{equation}
where $b_n = 1.9992 n - 0.3271$, $n$ is the \ser index, $(x_{light}, y_{light})$ are light coordinates, and $R$ is the half-light radius. In the modeling process, we again use the eccentricities as parameters instead of $q$ and the position angle $\phi$.}
We additionally model the light of the small nearby galaxy,
\ed{which also appears to be an elliptical galaxy, based on its color and morphology,} again using two elliptical \ser profiles.
\ed{Since the central regions of an elliptical galaxy are not always modeled well by \ser profiles \citep[e.g.,][]{shajib2021a}, a circular area with a radius of 2.5 pixels at its center (Figure~\ref{fig:what-to-mask}, dashed arrow) is masked.}
We do not model its mass.
We model the source using an elliptical \ser profile and \ed{shapelets} with $n_{max} = 6$ \citep{birrer2015a}.
We mask out the light for three very faint galaxies only visible in the \hst image (blue circles in Figure~\ref{fig:what-to-mask}).
We also mask out the small object embedded (in projection) in arc A (Figure~\ref{fig:what-to-mask}, solid arrow).\footnote{
We find that whether to include a mass model for this object \ed{makes} negligible difference for the main lens parameters, which is the goal of this modeling effort.
Determining whether it is a galaxy or a MW star may have implications for testing the detection of small perturbers in lensing \citep[e.g.,][]{vegetti2010a} but this is out of scope for this work and will be left to a future investigation.}
Our model consists of \nparams parameters, defined in Table~\ref{tab:parameters}. 

\clearpage


\begin{minipage}{\linewidth}
\makebox[\linewidth]{
\includegraphics[keepaspectratio=true,scale=0.75]{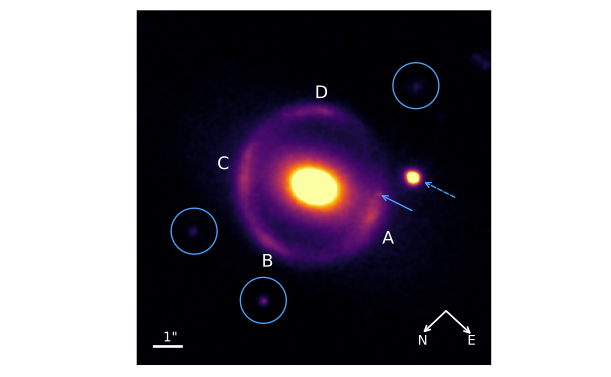}}
\captionof{figure}{
DESI-165.4754-06.0423 in \hst WFC3 F140W.
 The four lensed images are labeled clockwise as A, B, C and D.
 To model this system, we mask out the light from three faint objects (circles) and the small object (solid arrow) in arc A (in projection).
 We also mask out the central region of the small nearby galaxy with a circular mask of radius 2.5 pixel (dashed arrow).}
\label{fig:what-to-mask}
\end{minipage}


\begin{table}
\caption{Prior distribution used for lens modeling}
\begin{align*}
\text{Lens mass:} &
\begin{cases}
    \hfill \theta_E & \sim \exp\left(\mathcal{N}(\ln 2.5, 0.25\right) \\[1pt]
    \hfill \gamma & \sim \mathcal{TN}\left(2, 0.25, 1, 2.7\right) \\[1pt]
    \hfill \epsilon_{1}, \epsilon_{2} & \sim \mathcal{N}(0, 0.1) \\[1pt]
    \hfill x, y & \sim \mathcal{N}(0, 0.05) \\[1pt]
    \hfill \gamma_{ext, 1}, \gamma_{ext, 2} & \sim \mathcal{N}(0, 0.05)
\end{cases} \\
\text{Lens light:} &
\begin{cases}
    \hfill R_{\text{l}} & \ed{\sim \exp(\mathcal{N}(\ln 1, 0.15)) / \exp(\mathcal{N}({\ln 1, 0.15}))} \\[1pt]
    \hfill n_{\text{l}} & \ed{\sim \mathcal{U}(0.5, 10)/ \mathcal{U}(0.5, 10 )} \\[1pt]
    \hfill \epsilon_{\text{l},1}, \epsilon_{\text{l},2} & \ed{\sim \mathcal{N}(0, 0.15) / \mathcal{N}(0, 0.15)} \\[1pt]
    \hfill x_{\text{l}} & \ed{\sim \mathcal{N}(0, 0.05) / \mathcal{N}(0, 0.10)} \\[1pt]
    \hfill y_{\text{l}} & \ed{\sim \mathcal{N}(0, 0.05) / \mathcal{N}(0, 0.10)}
\end{cases} \\
\text{Source light:} &
\begin{cases}
    \hfill R_{\text{s}} & \sim \exp\left(\mathcal{N}(\ln 0.25, 0.15\right)) \\[1pt]
    \hfill n_{\text{s}} & \sim \mathcal{U}(0.5, 6) \\[1pt]
    \hfill \epsilon_{\text{s},1}, \epsilon_{\text{s},2} & \sim \mathcal{N}(0, 0.15) \\[1pt]
    \hfill x_{\text{s}}, y_{\text{s}} & \sim \mathcal{N}(0, 0.1) \\[1pt]
    \hfill \beta_{shp} & \sim \exp (\mathcal{N}(\ln 0.1, 0.1)) \\[1pt]
    \hfill x_{shp}, y_{shp} & \sim \mathcal{N}(0, 0.05)
\end{cases} \\
\text{Nearby galaxy:} &
\begin{cases}
    \hfill R_{\text{g}} & \ed{\sim \exp(\mathcal{N}(\ln 0.4, 0.2)) / \exp(\mathcal{N}({\ln 0.4, 0.2} ))} \\[1pt]
    \hfill n_{\text{g}} & \ed{\sim \mathcal{U}(0.5, 5)/\mathcal{U} (0.5, 5)} \\[1pt]
    \hfill \epsilon_{\text{g},1}, \epsilon_{\text{g},2} & \ed{\sim \mathcal{TN}\left(0, 0.15, -0.3, 0.3\right) / \mathcal{TN}\left(0, 0.15, -0.3, 0.3 \right)} \\[1pt]
    \hfill x_{\text{g}} & \ed{\sim \mathcal{TN}(3.7, 0.05, 3.55, 3.85) / \mathcal{TN}(3.7, 0.05, 3.55, 3.85)} \\[1pt]
    \hfill y_{\text{g}} & \ed{\sim \mathcal{TN}(0.25, 0.05, 0.1, 0.4) /\mathcal{TN}(0.25, 0.05, 0.1, 0.4)}
\end{cases}
\end{align*}
\label{tab:parameters}
{\small Notes --- 
The mass model consists of EPL for the lens mass profile and external shear ($\gamma_{ext}$). 
$\theta_E$ is the Einstein radius in arcsec, while $\gamma$ defines the slope of the EPL profile.
$x$ and $y$ are the mass center coordinates of the lens. 
$\gamma_{1, \text{ext}}$ \ed{and $\gamma_{2, \text{ext}}$} are the external shear components.
The parameters $\epsilon_1$ and $\epsilon_2$ are the lens mass eccentricities.
For the light model, subscripts $l$, $s$, $g$ indicate parameters belonging to the light profiles of the lens, the source, and the nearby galaxy, respectively. 
Thus $(\epsilon_{l,1}$, $\epsilon_{l,2})$, $(\epsilon_{s,1}$, $\epsilon_{s,2})$, $(\epsilon_{g,1}$, $\epsilon_{g,2})$ are the lens, source, and nearby galaxy light eccentricities, respectively.
For the light of the lens and the nearby galaxy, we use two elliptical S{\'e}rsic profiles each, with their prior distributions separated with a slash.
$R_l$, $R_s, R_g$ are the half-light radii, and $n_l$, $n_s, n_g$ are the S{\'e}rsic indices for the respective objects. 
$(x_l, y_l)$, $(x_s, y_s)$, $(x_g, y_g)$ describe their centers. 
The subscript $shp$ refers to the shapelets component for 
the source.
Finally, $\mathcal{U}(a, b)$ indicates a uniform distribution with support $[a, b]$, $\mathcal{N}(\mu, \sigma)$ is a Gaussian with mean $\mu$ and standard deviation $\sigma$, and $\mathcal{TN}(\mu, \sigma, a, b)$ is a truncated Gaussian with support $[a, b]$.}
\end{table}

We achieve excellent \ed{residuals} (Figure~\ref{fig:model}). 
The best-fit parameters are shown in \ed{Tables~\ref{tab:best-fit-params} and \ref{tab:best-fit-light-params}}.
The sampling results for the lens and external shear mass parameters are shown in Figure~\ref{fig:cornerplot}. 
For this system, the marginal distributions of the SVI posterior match the HMC sampling results for some parameter pairs as well as they do for the simulated systems presented in \citet[][their Figures~6 and A2]{gu2022a}, although the match is not as close for all pairs. Overall, the SVI posterior provides a good approximation to the HMC samples, demonstrating the efficacy of the SVI step.
We determine the Einstein radius of the lens to be $\theta_E = \thetaEfit$ 
and the slope of the power law mass profile to be $\gamma = \gammafit$.
The total mass within the critical curve is $1.7390_{-0.0043}^{+0.0046} \times 10^{12} \, M_{\odot}$. 
We use the larger $R_e$ from the two \ser models as the effective light radius for the lensing galaxy, \Refit, or \Refitkpc, \ed{adopting the cosmological parameters from \citet{planck2020a}: $H_0$ = 67.4 $\pm$ 0.5 km s$^{-1}$ Mpc$^{-1}$, $\Omega _{\Lambda}$ = 0.6847 $\pm$ 0.0073, and $\Omega _{M}$ = 0.315 $\pm$ 0.007}.
The best-fit model predicts magnifications of 
$25.0\pm 1.6$, $59.0\pm 1.6$, $51.1\pm 2.7$, and $35.43\pm 0.79$ for images A, B, C, and D, respectively (Figure~\ref{fig:what-to-mask}), computed as the average of the first and second methods in Appendix~\ref{sec:append-mag}.

As a further analysis on our best-fit $\gamma$, 
\ed{we explore regions higher than the prior for $\gamma$ in Table~\ref{tab:parameters} ($\mathcal{TN}(2,0.25,1,2.7)$) and demonstrate stability in the other parameters and the consistent tendency for $\gamma$ to approach the lower boundary.
We 
use two other priors for $\gamma$, while for all other parameters the prior remains unchanged.
}
One is a truncated Gaussian distribution centered around $2.0$ with boundaries at $1.5$ and $2.5$.
For the second, we use a uniform prior between $2.0$ and $3.0$.
In both cases, the corner plot shows consistent values with our best-fit model for all parameters except for $\gamma$, whose posterior is pushing against the lower boundary, $1.5$ and $2.0$, respectively. In addition, we observe a $42.13 \ \chi^2$ improvement for using the prior with the $1.5$ lower bound and a $551.87 \ \chi^2$ improvement for using the prior with the $2.0$ lower bound, favoring our best-fit model in both cases.



\begin{minipage}{\linewidth}
\vspace{5pt}
\makebox[\linewidth]{
\includegraphics[keepaspectratio=true,scale=0.40]{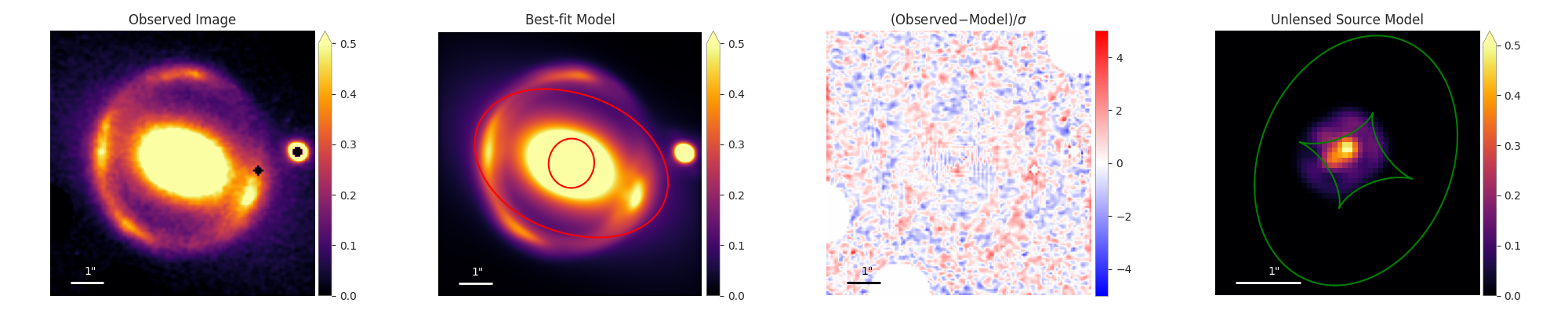}}
\captionof{figure}{From left to right, we show:
the observed \hst image (same orientation as in Figure \ref{fig:what-to-mask}), 
our best-fit model with critical curves in red, 
the reduced residual map,
and the reconstruction of the unlensed source with caustics in green. Notice the presence of an inner critical curve and caustic due to the fact that $\gamma<2$.
}
\label{fig:model}
\vspace{5pt}
\end{minipage}

Our sampling results are robust and statistically consistent using two widely employed metrics to measure the degree of convergence: 
the potential scale reduction factor (PSRF), \ed{often represented as \rhat} \citep{gelman1992a}, and the effective sample size (ESS). \ed{In cosmological studies, these two criteria have been widely adopted \citep[e.g.,][]{krolewski2021a, douspis2022a, mandel2022a, rubin2023a}.}
An $\hat{R}$ that is close to 1 and a large ESS indicate that convergence to the same stationary distribution has been achieved. 
\citet{gelman1992a} suggested that an appropriate condition is $\hat{R} < 1.2$.\footnote{Quoting from the \href{https://www.tensorflow.org/probability/api_docs/python/tfp/mcmc/potential_scale_reduction}{TensorFlow documentation}, ``Sometimes, $\rhat < 1.2$ is used to indicate approximate convergence." This implementation by TensorFlow, which we use, is approximately the square of the more recent (re)definition of \rhat---thus, using this definition, the equivalent threshold would be $\rhat < 1.1$ \citep{gelman2014a}.\label{fn:rhat}}


For our best-fit model, the \rhat values for all mass and light parameters of the lens and for all light parameters of the source are below 1.10. For the nearby galaxy light parameters, \rhat is less than 1.12 and thus, all parameters are well below the threshold of $\rhat < 1.2$.
As for ESS, $\mathcal{O}(10^3)$ is considered sufficiently high \citep[e.g.,][]{carpenter2017a, mandel2022a}. For our best-fit model, the minimum is 32,200, and the maximum, 40,000, demonstrating excellent sampling efficiency with very low autocorrelation \citep[see, e.g.,][]{gu2022a}. 

In addition to being robust, our modeling approach is also fast.
On a GPU node on the {\it Perlmutter} supercomputer at \ed{the National Energy Research Scientific Computing Center (NERSC)},\footnote{\url{https://www.nersc.gov/}} 
which has four A100 GPUs,
the total execution time is 3 hrs.\ and 13 min. (193 min.), 
with the MAP, SVI, and HMC steps taking 1 hr.\ 9 min. (69 min.), 1 hr.\ 52 min. (112 min.), and 12 min., respectively.
Considering that we take a fully forward modeling Bayesian approach with 41 model parameters for a highly nonlinear problem, 
this is remarkably fast.

\clearpage
\begin{deluxetable*}{lccccccc}
\tabletypesize{\scriptsize}
\tablecaption{Best-fit mass parameters for DESI-165.4754-06.0423.
\label{tab:best-fit-params}}
\renewcommand{\arraystretch}{1.4}
\tablehead{
    \colhead{$\theta_E$} &
    \colhead{$\gamma$} &
    \colhead{$\epsilon_1$} &
    \colhead{$\epsilon_2$} &
    \colhead{$x$} &
    \colhead{$y$} &
    \colhead{$\gamma_{ext, 1}$} &
    \colhead{$\gamma_{ext, 2}$} 
}
\startdata
$2.6463_{-0.0016}^{+0.0017}$& 
$1.372_{-0.022}^{+0.023}$&	
$0.1091_{-0.0020}^{+0.0020}$&$-0.1320_{-0.0020}^{+0.0020}$  & 
$0.0272_{-0.0022}^{+0.0023}$&	$-0.0018_{-0.0017}^{+0.0017}$ &
$0.0657_{-0.0024}^{+0.0024}$&	
$-0.0939_{-0.0022}^{+0.0022}$ \\
\enddata
\end{deluxetable*}

\begin{deluxetable}{l|rr|r|rr}
\tablecaption{Best-fit light parameters for DESI-\ed{165.4754-06.0423.}
\label{tab:best-fit-light-params}
}
\renewcommand{\arraystretch}{1.4}
\setlength{\tabcolsep}{10pt}
\tablehead{
    \colhead{Parameter} & \multicolumn{2}{c}{Lens light} & \colhead{Source light} & \multicolumn{2}{c}{Nearby galaxy light} 
}
\startdata
$\bm{R_e}$ &
$0.4327_{-0.0050}^{+0.0051},$&$1.888_{-0.017}^{+0.017}$&$0.614^{+0.038}_{-0.036}$&     $0.1353_{-0.0025}^{+0.0021},$&$0.227_{-0.065}^{+0.066}$\\[4pt]
$\bm{n}$ &               
$0.925_{-0.014}^{+0.014},$&$1.262_{-0.033}^{+0.035}$&$0.5050^{+0.0082}_{-0.0037}$&
$0.523_{-0.017}^{+0.032},$&$3.86_{-1.22}^{+0.81}$\\[4pt]
$\bm{\epsilon_{1}}$ &    
$0.1350_{-0.0022}^{+0.0020},$&$0.1452_{-0.0012}^{+0.0012}$&$-0.058^{+0.025}_{-0.026}$&
$0.0118_{-0.0089}^{+0.0087},$&$0.144_{-0.074}^{+0.059}$\\[4pt]
$\bm{\epsilon_{2}}$ &    
$-0.0577_{-0.0038}^{+0.0039},$&$-0.1832_{-0.0013}^{+0.0013}$&$0.192^{+0.023}_{-0.023}$& 
$-0.1124_{-0.0102}^{+0.0080},$&$0.264_{-0.055}^{+0.026}$\\[4pt]
$\bm{x}$ &
$0.0042_{-0.0014}^{+0.0013},$&$0.0437_{-0.0023}^{+0.0024}$&$0.137^{+0.013}_{-0.012}$& 
$3.6100_{-0.0039}^{+0.0023},$&$3.7347_{-0.0079}^{+0.0062}$\\[4pt]
$\bm{y}$ &
$0.0448_{-0.0012}^{+0.0013},$&$-0.0053_{-0.0018}^{+0.0018}$&$0.044^{+0.012}_{-0.012}$&
$0.3704_{-0.0011}^{+0.0014},$&$0.3580_{-0.0058}^{+0.0043}$\\[4pt]      
$\bm{\beta_{shp}}$ &
- & - & $0.0505_{-0.0030}^{+0.0031}$ & -  & - \\[4pt]
$\bm{x_{shp}}$ & - & - &$-0.0249_{-0.0040}^{+0.0039}$& - & - \\[4pt]
$\bm{y_{shp}}$ & - & - &$0.0354_{-0.0025}^{+0.0025}$& -  & - \\  
\enddata

\vspace{3pt}
\ed{\noindent \hspace{0.5em}{\small Notes ---
We model the light of both the lens and nearby galaxy with two elliptical Sérsic profiles each,
with the best-fit parameters shown in the second and last main columns, respectively.}}
\end{deluxetable}

\begin{minipage}{\linewidth}
\makebox[\linewidth]{
  \includegraphics[keepaspectratio=true,scale=0.35]
  {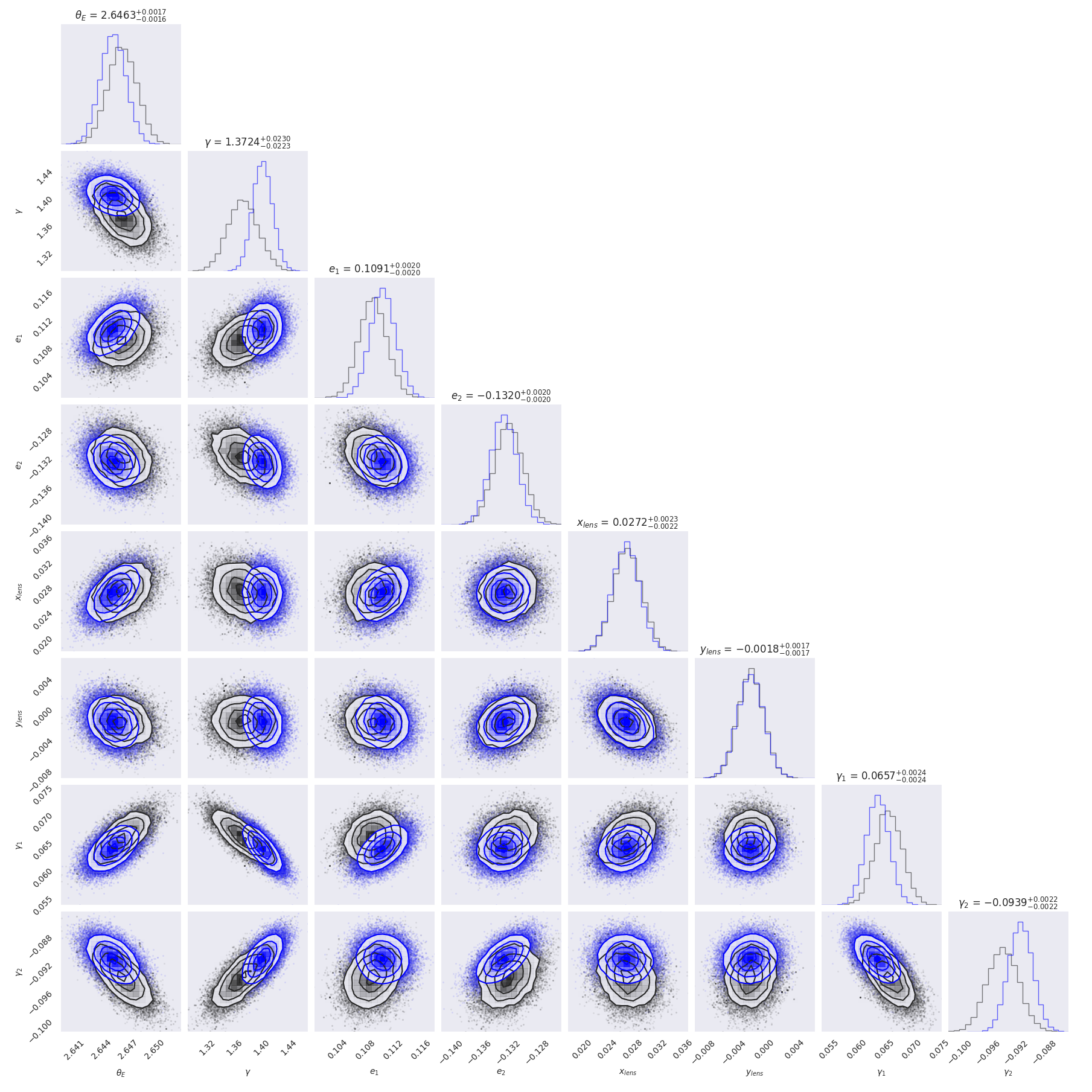}}
\captionof{figure}{The corner plot showing our sampling results for the lens and external shear mass parameters\ed{,
with SVI and HMC samples shown in blue and black, respectively.}
}

\label{fig:cornerplot}
\end{minipage}

\section{Discussion}\label{sec:discussion}
\ed{The accurate measurements of the density slopes of lensing galaxies, which are typically elliptical galaxies, have significant implications for time-delay \ho constraints \citep[e.g.,][]{birrer2020a}
and are crucial for determining other cosmological parameters using a large number of static lenses \citep[e.g.,][]{li2024a}.
Such measurements for lensing systems over a wide range of redshifts also makes it possible to study
the structural evolution of massive elliptical galaxies \citep[e.g.,][]{sahu2024a}.}

Numerical simulations of cold dark matter (CDM) halos predict mass profiles with a logarithmic slope approaching $\gamma = 1$ and 3 
at small and large radii, respectively \citep[e.g.,][NFW]{navarro1996a}.
In cluster strong lenses, low $\gamma$ values near the cluster center certainly \ed{have} been measured \citep[e.g.,][at 1.26]{broadhurst2000a}.
For galaxy-scale lenses, with significant baryon  contribution (typically stellar in elliptical galaxies) to the mass, 
the picture is more complicated.
Nevertheless, low $\gamma$ values around 1.4 \ed{have} also been reported, 
for five systems from the BELLS and SL2S samples \citep{LiR2018a} and one from the LSD sample \citep{treu2004a}. 
There is an even larger number of systems (three from SLACS and six from BELLS) for which the $\gamma$ value is consistent with our best-fit value for DESI-165.4754-06.0423 to within 2$\sigma$.
As mentioned earlier, this system appears to be at the center of a galaxy group, 
and such systems have not been as extensively studied compared with galaxy-scale strong lenses that typically have smaller Einstein radii.
In addition, the previous analyses on the LSD, SLACS, BELLS, and SL2S samples were not \ed{performed} with a fully forward modeling approach\ed{, nor have convergence metrics been applied to those results}.
Thus applying this approach
consistently to a large number of systems with satisfactory sampling convergence for each system
(see \S\,\ref{sec:lens-model} and the paragraph below) 
can shed more light on the mass slope question.
Finally, for this system,
comparison with mass profile modeling based on well-measured velocity dispersion, 
especially from spatially resolved stellar dynamics \ed{\citep[e.g.,][]{treu2004a, barnabe2009a,  shajib2023a, turner2024a}} will be very interesting and is left for a future investigation.

Given the covariance that exists between the fitting parameters of a model, 
to properly estimate the uncertainties of these parameters, 
it is best to perform full forward modeling.
However, to take this approach for strong lensing is challenging given that it is inherently a highly nonlinear problem 
(in fact, the solution is near \ed{a mathematical singularity}, the critical curve).
Using high resolution data (e.g., \hst) for lens modeling would require a high dimensional parameter space and, if the Einstein radius is large (which is the case for this system), a large cutout---both will present even greater challenges.
Thanks to modern GPUs and automatic differentiation, 
this became possible in recent years and we developed \gigal.
In this pipeline, we also assess convergence for statistical inference, 
using the metrics of \rhat and ESS.
We show that for this system we have achieved excellent convergence.
Thus this first attempt has yielded very encouraging results.

\ed{In the event of a SN in the source galaxy,
the precise time-delays will depend on the location of the SN. From using different test locations, 
the arrival time offset between the earliest and latest images is $\gtrsim 40$ days, which would be ideal for \ho measurements.
We will determine the stellar mass and star formation rates to estimate the SN rates, based on the photometric and spectroscopic data for this and other systems in follow-up analyses.}
\ed{Finally, from this modeling effort, we do not see evidence for dark matter substructure or line-of-sight low-mass halos in this system. 
In future publications, we will report the modeling results for other systems in this program, including the detection status of substructure/low-mass LOS halos.}
\section{Conclusion}\label{sec:conclusion}

In this paper---Paper I of the \ed{\projname} series---we present our \HST SNAP program (GO-15867, PI: Huang) for confirming strong gravitational lens candidates found using Residual Neural Networks in the DESI Legacy Imaging Surveys.
\ed{It is the first \hst program to follow up on strong lens candidates found using machine learning in imaging surveys.}
All 51 systems observed by \hst were confirmed to be lenses.
To our knowledge, this is the first time an \hst~SNAP program for strong lens confirmation has achieved a 100\% success rate.

Spectroscopic observations of these systems are ongoing through DESI (Paper~II in this series; Huang, Inchausti et al., in prep) and on Keck NIRES (Paper~III in this series, Agarwal et al. in prep).

We have applied a \emph{fully} forward modeling Bayesian approach (\gigal) using \emph{multiple} GPUs, for the first time in both regards, to a real gravitational
lensing system with \hst (or any high resolution) data, DESI-165.4754-06.0423.
The Einstein radius of the lens is $\theta_{E} =\thetaEfit$ and the slope of the power law mass profile is $\gamma = \gammafit$.
We report excellent sampling results: 
for all parameters 
the \ed{$\hat{R}$ is well-below the threshold of 1.2 (based on the TensorFlow definition and implementation; see Footnote\footref{fn:rhat}})
and the effective sample size (ESS), greater than 37,000.
Our model is also fast.
For such a complex model (\nparams parameters) for a highly nonlinear problem,
with a large \hst image size ($128^2$ pixels),
the time it takes to run \gigal is $\exetime$ on four A100 GPU on one GPU node on the \emph{Perlmutter} supercomputer at \ed{NERSC}. 
With further improvement in the hardware and software almost a certainty, 
this concretely demonstrates a promising future of modeling  many more strong lensing systems with \hst data, or observed with comparable or superior image resolution and depth (e.g.,
\emph{Euclid}, the \JWST, and the \RST), with speed and statistical rigor.

\ed{
Recently two cluster lensed SNe have been used to determine \ho \citep{kelly2023a, pascale2024a}, with a measurement from a third system possible \citep{pierel2023a}.
For the two resolved galaxy-scale strongly lensed SNe discovered so far, the Einstein radii (\tE), and therefore time-delays ($< 1$~day), are too small for this to be possible \citep{mortsel2020a, goobar2023a, pierel2023a}.
However, with 
the much larger sample of galaxy-scale strong lenses already observed with high resolution imaging and modeled, with a time-varying source \citep[e.g.,][]{suyu2010a, suyu2013a, wong2019a, shajib2023a, schmidt2023a} or without \citep[e.g.,][]{bolton2008a, shu2016a},
there are still significant advantages of using galaxy-scale lensed SNe for \ho measurement: the modeling is simpler and the systematics are well-understood \citep[e.g.,][]{kochanek2020a, birrer2020a, shajib2023a}.
From the ongoing work of modeling the galaxy-scale strong lenses in this program beyond the system presented in this work, 
we will measure the lens mass slope over a wide redshift range, providing robust prior for not only \ho measurements, but the determination of other cosmological parameters \citep[e.g.,][]{li2024a}.
In addition, building on the experience we have gained using \gigal from modeling systems in this sample, we have extended this framework to lensed point sources 
(e.g., SNe; S.~Baltasar, N.~Ratier-Werbin \& X.~Huang in prep).
Finally, this framework has also been extended to model group/cluster scale lenses \citep{urcelay2025a}, 
which we will apply to the group/cluster strong lenses in this sample.
}

\newpage
\section*{Acknowledgement}\label{sec:acknowledgement}
We thank Peter Harrington, Nestor Demeure, Steve Farrell, and Rollin Thomas at the National Energy Scientific Computating Center (NERSC) for their consultation and advice.
This research used resources of \ed{NERSC}, a U.S. Department of Energy Office of Science User Facility operated under Contract No. DE-AC02-05CH11231 and the Computational HEP program in The Department of Energy's Science Office of High Energy Physics provided resources through the ``Cosmology Data Repository" project (Grant \#KA2401022).
X.H. acknowledges the
University of San Francisco Faculty Development Fund.
Support for HST program 15867 was provided by NASA
through a grant from the Space Telescope Science Institute,
which is operated by the Association of Universities for
Research in Astronomy, Inc., under NASA contract NAS
5-26555.
A.D.'s research is supported by National Science Foundation's National Optical-Infrared Astronomy Research Laboratory, 
which is operated by the Association of Universities for Research in Astronomy (AURA) under cooperative agreement with the National Science Foundation.

This paper is based on observations at Cerro Tololo Inter-American Observatory, National Optical
Astronomy Observatory (NOAO Prop. ID: 2014B-0404; co-PIs: D. J. Schlegel and A. Dey), which is operated by the Association of
Universities for Research in Astronomy (AURA) under a cooperative agreement with the
National Science Foundation.

This project used data obtained with the Dark Energy Camera (DECam),
which was constructed by the Dark Energy Survey (DES) collaboration.
Funding for the DES Projects has been provided by 
the U.S. Department of Energy, 
the U.S. National Science Foundation, 
the Ministry of Science and Education of Spain, 
the Science and Technology Facilities Council of the United Kingdom, 
the Higher Education Funding Council for England, 
the National Center for Supercomputing Applications at the University of Illinois at Urbana-Champaign, 
the Kavli Institute of Cosmological Physics at the University of Chicago, 
the Center for Cosmology and Astro-Particle Physics at the Ohio State University, 
the Mitchell Institute for Fundamental Physics and Astronomy at Texas A\&M University, 
Financiadora de Estudos e Projetos, Funda{\c c}{\~a}o Carlos Chagas Filho de Amparo {\`a} Pesquisa do Estado do Rio de Janeiro, 
Conselho Nacional de Desenvolvimento Cient{\'i}fico e Tecnol{\'o}gico and the Minist{\'e}rio da Ci{\^e}ncia, Tecnologia e Inovac{\~a}o, 
the Deutsche Forschungsgemeinschaft, 
and the Collaborating Institutions in the Dark Energy Survey.
The Collaborating Institutions are 
Argonne National Laboratory, 
the University of California at Santa Cruz, 
the University of Cambridge, 
Centro de Investigaciones En{\'e}rgeticas, Medioambientales y Tecnol{\'o}gicas-Madrid, 
the University of Chicago, 
University College London, 
the DES-Brazil Consortium, 
the University of Edinburgh, 
the Eidgen{\"o}ssische Technische Hoch\-schule (ETH) Z{\"u}rich, 
Fermi National Accelerator Laboratory, 
the University of Illinois at Urbana-Champaign, 
the Institut de Ci{\`e}ncies de l'Espai (IEEC/CSIC), 
the Institut de F{\'i}sica d'Altes Energies, 
Lawrence Berkeley National Laboratory, 
the Ludwig-Maximilians Universit{\"a}t M{\"u}nchen and the associated Excellence Cluster Universe, 
the University of Michigan, 
{the} National Optical Astronomy Observatory, 
the University of Nottingham, 
the Ohio State University, 
the OzDES Membership Consortium
the University of Pennsylvania, 
the University of Portsmouth, 
SLAC National Accelerator Laboratory, 
Stanford University, 
the University of Sussex, 
and Texas A\&M University.

\ed{This research used data obtained with the Dark Energy Spectroscopic Instrument (DESI). DESI construction and operations is managed by the Lawrence Berkeley National Laboratory. This material is based upon work supported by the U.S. Department of Energy, Office of Science, Office of High-Energy Physics, under Contract No. DE–AC02–05CH11231, and by the National Energy Research Scientific Computing Center, a DOE Office of Science User Facility under the same contract. Additional support for DESI was provided by the U.S. National Science Foundation (NSF), Division of Astronomical Sciences under Contract No. AST-0950945 to the NSF’s National Optical-Infrared Astronomy Research Laboratory; the Science and Technology Facilities Council of the United Kingdom; the Gordon and Betty Moore Foundation; the Heising-Simons Foundation; the French Alternative Energies and Atomic Energy Commission (CEA); the National Council of Science and Technology of Mexico (CONACYT); the Ministry of Science and Innovation of Spain (MICINN), and by the DESI Member Institutions: www.desi.lbl.gov/collaborating-institutions. The DESI collaboration is honored to be permitted to conduct scientific research on Iolkam Du’ag (Kitt Peak), a mountain with particular significance to the Tohono O’odham Nation. Any opinions, findings, and conclusions or recommendations expressed in this material are those of the author(s) and do not necessarily reflect the views of the U.S. National Science Foundation, the U.S. Department of Energy, or any of the listed funding agencies.}


\software{
    \texttt{GIGA-Lens} \citep{gu2022a},
    TensorFlow \citep{TensorFlow},
    TensorFlow Probability \citep{dillon2017a}, 
    JAX \citep{bradbury2018a}, 
    Optax \citep{optax2020},
    \texttt{Lenstronomy} \citep{birrer2018a},
    Matplotlib \citep{hunter2007a},
    photutils \citep{bradley2023a},
    seaborn \citep{waskom2021a},
    corner.py \citep{foreman2016a},
    NumPy \citep{harris2020a},
    Astropy \citep{astropy22a}
}

\bibliographystyle{aasjournal}
\bibliography{dustarchive}

\appendix

\section{Magnification} \label{sec:append-mag}
This appendix describes the magnification calculations for the lensed arcs. 
We will show three independent methods that provide consistent results with each other.

In the first method, we choose a reasonable luminosity threshold for the source light. 
The pixels in the source plane and the lens plane with values exceeding the threshold are selected, with both images being non-PSF convolved. 
This threshold is chosen in a manner that allows each arc to be isolated from the others. 
The ratios between the number of pixels of each arc and the number of pixels of the source, above this threshold, will be the magnifications. 
In practice, an important adjustment needs be made in order to achieve accurate results.
Each flux value in the image plane comes from a corresponding pixel in the source plane. 
\gigal ensures this correspondence by ray-tracing each pixel of the image plane to the source plane.
It then assigns the pixel in the image plane with the value of the corresponding pixel on the source plane. 
However, for magnified images, using a grid in the source plane that is the same as the lens plane can lead to inaccuracy in the magnification calculation.
Such a grid proves to be too coarse and thus
leads to an underestimation of the magnification.
We use the peak pixel values in the source and image planes as a diagnostic.
We find that the source plane grid is sufficiently fine if a correspondence between the peak pixel values in the image and source planes can be established.
We thus proceed as follows. 
Let $\delta_I$ and $\delta_S$ be the pixel sizes, measured in arcseconds, of the image and source planes, respectively, with a sufficiently small $\delta_S$ (for typical lensed images, which are magnified, $\delta_I > \delta_S$).
In the case of DESI-165.4754-06.0423, we use  $\delta_I = 0.065''$ and $\delta_S = 0.02''$.
We can now count pixels and compute the ratio for each arc. 
We account for the pixel size difference between the planes
by multiplying the ratio by the factor $(\delta_I/\delta_S)^2$. This is the magnification.
This method is straightforward, but it requires determining a reasonable grid resolution for the source plane. 


We now introduce a second method that does not require that we choose an appropriate resolution in the source plane.
For each pixel that exceeds a chosen threshold (identical to the one used in the first method) within a non-PSF-convolved arc, we ray-trace the center of that pixel to the source plane.
We then compute the area of the convex hull that encloses these delensed points.  
The magnification is the ratio between the area of each arc and the source area calculated this way,
as long as source structure does not have concave features and the convex hull results in a reasonable enclosure.\footnote{\citet{zhang2023a} uses convex hull in the \emph{image} plane for magnification calculations.
Given that a lensed arc can have significant curvature, the resulting enclosure can sometimes cover a larger area than the arc, and lead to inaccurate magnification.}

As a final check, we apply the point-wise magnification function from \texttt{Lenstronomy} to our model. The mean values of all points within each arc are also consistent with the magnification for each arc calculated using the first two methods. We present the magnification results for these three methods in Table~\ref{tab:magnifications}.



In the literature, magnifications are typically calculated using one method. 
Here we take three reasonable approaches, and we note that the differences in the magnifications from using these different methods are greater than the statistical uncertainties. Thus, we regard the differences between the methods as an estimate for the uncertainty. For all three methods, the magnification for all four lensed images are in agreement to $\lesssim 10\%$. 
We take the averages of the first two methods as the magnifications and their differences divided by two as the uncertainty. This is what is reported in the main text.
We treat the third method as a ``sanity check''.

\begin{deluxetable*}{lccccc}
\tabletypesize{\footnotesize}
\tablecaption{Magnification calculations for DESI-165.4754-06.0423.
\label{tab:magnifications}}
\renewcommand{\arraystretch}{1.4}
\tablehead{
    \colhead{Method} &
    \colhead{$A$} &
    \colhead{$B$} &
    \colhead{$C$} &
    \colhead{$D$} &
    \colhead{Total}
}
\startdata
$1^{\text{st}}$&$26.65$&$57.33$&$53.81$&
$36.21$&$174.00$ \\
$2^{\text{nd}} $&$23.41$&$60.61$&$48.43$& 
$34.64$&$167.09$ \\
$3^{\text{rd}} $&$24.83$&$57.15$&$53.06$& 
$34.95$&$169.99$ \\
\enddata
\tablecomments{Our magnification calculations, using three approaches, for images A, B, C, D (see Figure~\ref{fig:what-to-mask}), and the total magnification, respectively.
}
\end{deluxetable*}

\end{document}